\documentclass[%
 reprint,
 amsmath,amssymb,
 aps,
longbibliography, pra]{revtex4-2}

\usepackage{times}
\usepackage[table]{xcolor}
\usepackage{graphicx} 
\usepackage{ulem}
\usepackage{bm, amsmath, amssymb, mathtools}

\usepackage{soul}
\usepackage{siunitx}

\bgroup
\catcode`@=11
\gdef\getfontsize{\f@size pt}
\egroup

\makeatletter
\edef\textFontName{\fontname\csname
  \f@encoding/\f@family/\f@series/\f@shape/\f@size\endcsname}

\makeatother

\begin{document}

\preprint{APS/123-QED}

\title{Mixing of counterpropagating signals in a traveling-wave Josephson device}

\author
{M. Praquin,$^{1}$  A. Giraudo,$^{1}$ V. Lienhard,$^{1,2}$ T. Bouwakdh,$^{1}$ A. Vanselow,$^{1}$ Z. Leghtas,$^{1}$ 
P. Campagne-Ibarcq$^{1}$}
\affiliation{$^1$ Laboratoire de Physique de l’Ecole Normale Supérieure, Mines Paris-PSL,
Inria, ENS-PSL, Université PSL, CNRS, Sorbonne Université, Paris, France\\
$^2$ Present address: Ohmori Group, Okazaki, Japan}

\date{\today}

\vspace{1cm}

\begin{abstract}
Light waves do not interact in vacuum, but may mix through various parametric processes when traveling in a nonlinear medium. In particular, a high-amplitude wave can be leveraged to frequency convert a low-amplitude signal, as long as the overall energy and momentum of interacting photons are conserved. These conditions are typically met when all waves propagate in the medium with comparable phase velocity along a particular axis. In this work, we investigate an alternative scheme  by which   an input microwave  signal propagating  along a 1-dimensional Josephson metamaterial is converted to an output wave  propagating in the opposite direction. The interaction is mediated by a pump wave propagating at low phase velocity. In this regime, the input signal is exponentially attenuated as it travels down the device. We exploit this process to implement a robust on-chip microwave isolator that can be reconfigured  into a reciprocal and tunable coupler. The device mode of operation is selected \textit{in situ}, along with its  working frequency over a wide microwave range. We measure an isolation over 5~dB in the  5-8.5~GHz range and over 10~dB in the 7-8.5~GHz range on a typical bandwidth of 200~MHz. Substantial margin for improvement exists through design optimization and by reducing fabrication disorder, opening new avenues for microwave routing and processing in superconducting circuits.

\end{abstract}

\maketitle 

\section{Introduction}
\label{sec:introduction}
Parametric processes are potent and versatile tools in  quantum technologies, enabling  signal amplification~\cite{Hansryd2002,HoEom2012}, harmonic generation~\cite{Efimov2003}, controlled interaction between standing modes~\cite{Lecoq2017}, non-reciprocal signal transmission~\cite{Kamal2011, Abdo2013} and entanglement generation~\cite{Kwiat1995}, both in the optical and microwave domains. In these processes, input waves mix in a non linear medium to produce output waves such that energy is conserved:
\begin{equation}
\sum_{i=1}^N \epsilon_i \omega_i=0
\label{eq:energy}
\end{equation}
where $\epsilon_i=1$ for input waves $\epsilon_i=-1$ for output waves, $\omega_i$ is the frequency of each wave, and the  number of involved waves $N$ matches the order of the wave-mixing process. Since the process rate scales as the product of wave amplitudes, a strong \textit{pump} wave (P) is typically employed to enhance interactions between weak \textit{signal} (S) and \textit{idler} (I) waves. When $N=4$---which is the case for the $\chi^{(3)}$ medium considered in this work---a pump at frequency $\omega_P=(\omega_S+\omega_I)/2$ is typically employed to activate a two-mode squeezing interaction between the signal and idler waves and  amplify the input signal close to the quantum limit. In another mode of operation, a pump at frequency $\omega_P=|\omega_S-\omega_I|/2$ activates a beam-splitting interaction between the waves, converting an input signal to an output idler and vice-versa. \\

In the first experimental implementations of these ideas, the non linear medium was embedded in a high quality factor resonator in order to enhance interactions~\cite{Yamamoto2008, Bergal2010, Roch2012, Mutus2013, Mutus2014, Eichler2014, Roy2015, Sivak2019}, at the expense of limiting the process working bandwidth.  Alternatively, one may send waves through a long section of non linear medium, which is known as a traveling wave parametric process~\cite{Cullen1958, Tien1958, Yaakobi2013}. For the interactions along this line to add up constructively, phase-matching must be enforced which results in a second conservation law on momentum
\begin{equation}
\sum_{i=1}^N \epsilon_i k_i=0,
\label{eq:momentum}
\end{equation}
where $k_i$ is the wavevector corresponding to each wave.  {While waves naturally propagate at the same speed in a dispersionless linear medium---trivially linking equations \eqref{eq:energy} and \eqref{eq:momentum} since $\omega_{i}=v_{i}\,k_{i}$---phase-matching is difficult to satisfy in a nonlinear medium due to the distinct wavevector shifts induced by the pump on the various waves.} Thus, the onset of traveling wave parametric amplifiers (TWPAs) allowed for broadband and high dynamic range amplification, at the cost of increased design complexity~\cite{Castellanos2008, Bockstiegel2014, OBrien2014, Macklin2015, White2015, Bell2015, Vissers2016, Zorin2016, Adamyan2016, Chaudhuri2017, Zhang2017, Ranzani2018, Miano2019, Planat2020, Goldstein2020, Malnou2021, Ranadive2022, Peng2022, Qiu2023,wang2025high,chang2025josephson,kow2025traveling} to enforce phase-matching.\\

\begin{figure}
    \centering
    \includegraphics{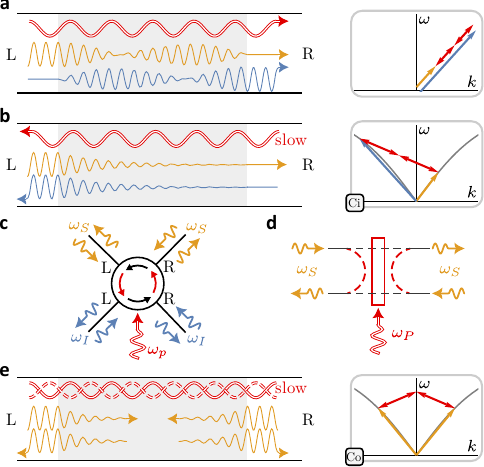}
    \caption{{\bf Traveling wave parametric conversion.} {\bf a,} In a regular traveling wave device, signal (orange), idler (blue) and pump waves (red) co-propagate   in a $\chi^{(3)}$ non-linear medium (gray shaded area), where they mix to coherently convert back and forth signal and idler. Signal isolation is only perfect if the conversion is total at the right port of the device (R).  Right panel : in energy-momentum space, waves are represented by arrows arranged in a  {closed} form when the process is phase-matched. If all waves propagate with same velocity (aligned arrows), the device is broadband. {\bf b,} In our device,  signal and idler waves propagate  in opposite directions such that conversion  attenuates  the signal exponentially. The process is phase-matched by lowering the pump velocity. Right panel : phase-matching is possible at any given signal frequency, irrespective of the exact dispersion relation of the medium (curved gray line), by adjusting the pump frequency.  {\bf c,} Signal waves entering from the device left port (L) and idler waves entering from the right port (R) are  reflected and frequency converted (red curved arrows) following the process pictured in (b). The same waves entering from the opposite port are transmitted as conversion is not phase-matched in that case, implementing a 4-port circulator (2 physical ports and 2 frequencies of interest). {\bf d,} Our device is reconfigurable into a reciprocal  coupler whose transparency is set by the pump amplitude. {\bf e,} This is achieved by feeding the device with pump waves from both ports, generating a standing wave in the nonlinear medium.  }
    \label{fig0}
\end{figure}

Traveling wave parametric converters (TWPCs) ~\cite{Ranzani2017,malnou2024traveling,ranadive2024traveling}, capable of converting  waves from signal to idler frequency, have  attracted interest as they are non reciprocal devices: if one  {reverses} the direction of propagation of the signal and idler waves, their propagation velocities $v_{S,I}$ become negative and the now unmatched conversion process is inhibited. Implementing a robust non-reciprocal element with high enough bandwidth and isolation to efficiently route microwave signals is a long standing goal of circuitQED~\cite{Kamal2014, Metelmann2015, Sliwa2015, Chapman2017, wu2019isolating, abdo2019active, zhang2021magnetic, abdo2021high, Naghiloo2021, Kwende2023, navarathna2023,beck2023wideband, Cao2024, kumar2024fanoenhanced}. However, TWPCs suffer from the same design complexity as TWPAs and from limited isolation as the non linear medium length and pump amplitude need be precisely adjusted for the conversion from signal to idler to be complete (see Fig~\ref{fig0}a).\\

In this work, we introduce a different regime of traveling wave conversion where the signal and idler propagate in opposite directions. Phase-matched conversion is achieved in this configuration by lowering significantly the pump wave velocity $|v_P|\ll |v_{I,S}|$,  {in a configuration analog to Stimulated Brillouin Scattering in optics~\cite{boyd2008nonlinear}\footnote{In stimulated Brillouin scattering, the low velocity wave is acoustic and interacts with light through electrostriction. }}, (see Fig~\ref{fig0}b). This change of direction of propagation upon conversion drastically changes the waves dynamics : as the idler escapes backward, it cannot be back converted to signal and the signal amplitude decreases exponentially along the line. Moreover, the device may be reconfigured to  enable controllable reciprocal transmission with high on/off ratio. We now detail these two modes of operation.\\

\textbf{On-chip circulation}: in this non-reciprocal process, conversion of a signal to an idler photon is enabled by the absorption of 2 photons from a pump wave co-propagating with the idler. The reverse conversion process from idler to signal leads to the emission of 2 photons in the pump wave. The corresponding phase matching diagram is presented in Fig.~\ref{fig0}b. From relations \ref{eq:energy} and \ref{eq:momentum}, we find the process to be phase-matched for:
\begin{equation}
\label{eq:signal-freq}
\omega_S=2\omega_P\frac{1/v_I~-~1/v_P}{1/v_S~-~1/v_I} \quad \omega_I=\omega_S+2\omega_P
\end{equation}
where we recall that $v_I, v_P<0$. Crucially, if one reverses the direction of propagation of both the signal and idler waves, this process is no longer phase-matched and waves propagate unaffected along the  transmission line. The device thus circulates signals through 2 frequencies of interest ($\omega_S$ and $\omega_I$) at either end of the transmission line, implementing an effective 4-port circulator (see Fig~\ref{fig0}c).\\

\textbf{Tunable coupling}: in this configuration, pump waves of same frequency are applied in both directions. A signal photon may be reflected back to an idler photon \textit{of same frequency} $\omega_S$ by absorbing a photon from one of the pump waves and emitting it back in the other (see Fig~\ref{fig0}d-e).
 We find that the process is phase-matched for :
\begin{equation}
\label{eq:coupler-freq}
\omega_S=\omega_P | \frac{v_S}{v_P} | \quad \omega_I=\omega_S
\end{equation}
Since this process is stimulated by pump photons, adjusting the pump amplitudes allows one to continuously tune the effective transmission and reflection coefficients of the line, implementing a tunable coupler.\\

We note that phase-matching of the target process is straightforward in both  configurations as no assumptions are imposed on the relative phase velocities of the propagation modes. It is however exactly verified for a precise set of wave frequencies only. In practice, the operating band of the device (typically of order  {200}~MHz in our experiment, see section~\ref{sec:amp-dependency} and supplemental~\ref{sec:exp-bandwidth}) is set by the pump amplitude. Its center frequency is tunable over a wide frequency range by adjusting the pump frequency.

\begin{figure}

\includegraphics{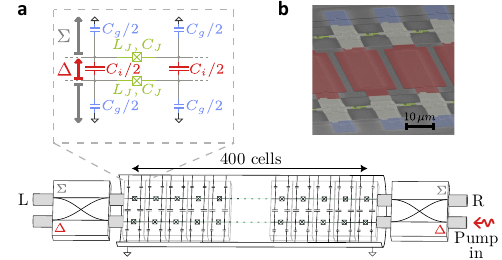}
    \caption{\label{fig1}    {\bf Design and fabrication. a,} Our device consists of 400 identical cells arranged in series, forming two  coplanar superconducting electrodes doped with Josephson junctions (green squared crosses). In each cell, the electrodes are capacitively coupled to each other (red capacitors) and to the ground reference plane (blue capacitors). Given its symmetry, this transmission line embeds two propagation modes with common ($\Sigma$) or differential ($\Delta$) phase across the electrodes, both supporting currents through the junctions. Each mode is selectively addressed via hybrid couplers connected at each end of the line. The pump propagates on the  $\Delta$ mode and is slowed down by the large inner capacitors. The probe signal propagates on the faster $\Sigma$ mode.  {\bf b,} False color scanning electron microscope photography of the device, see supplemental~\ref{sec:fab} for details of the fabrication process. }
\end{figure}
\section{Device design and characterization}

\begin{figure}

\includegraphics{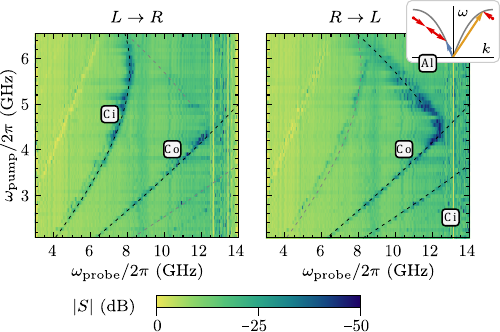}
    \caption{\label{fig2}{\bf Forward and backward transmissions} of a probe applied to the $\Sigma$ mode  {(referenced at the $\Sigma$ ports of the hybrid couplers in Fig.~\ref{fig1}a)} in presence of a pump traveling backward on the $\Delta$ mode, against pump and probe frequencies. Gaps in transmission (dark blue) reveal phase-matched processes identified as the circulation (Ci) and tunable coupling (Co) processes of Fig~\ref{fig0}b-e, and a process (Al) appearing close to the line cut-off frequency, resulting from aliasing. The  location of the gaps is  captured  with the lumped element model pictured in Fig~\ref{fig1}a (dashed lines), see supplemental~\ref{sec:bands}. The depth of the Ci and Al gaps is different  in each direction of propagation, showing non-reciprocity. The Co gap should not be visible if the pump was purely unidirectional, indicating internal pump reflection. }
\end{figure}
We design and fabricate a microwave transmission line made of lumped circuit elements and embedding two modes of propagation with different velocities (see Fig~\ref{fig1}). Wave-mixing is enabled by Josephson junctions supporting currents from both modes. Note that mixing of counterpropagating signals could also be achieved in a single-mode transmission line if its dispersion relation were properly engineered to slow down the pump wave. Beyond its simple design and  {versatility}, our two-mode architecture allows for straightforward separation of the high-amplitude pump wave  from the signals to be processed.\\

In detail, the device is composed of 400 identical cells arranged in series forming two inner electrodes and a ground reference electrode. In each cell, the two inner electrodes are interrupted by two Josephson junctions (inductance $L_J=0.94$~nH), shunted to ground by two identical  parallel-plate  capacitors (capacitance $C_g=0.13$~pF) and connected to each other  by a large capacitor ($C_i=0.57$~pF). The line symmetry ensures that the common  and differential excitations of the inner electrodes (labeled $\Sigma$ and $\Delta$ respectively) are eigenmodes of propagation. The large inner capacitor loads the $\Delta$ propagation mode---supporting the pump wave---while not affecting the $\Sigma$ mode---supporting the signal and idler waves of higher velocity. For low-frequency  {and low-amplitude} waves, the mode velocities read 
\begin{align}
\begin{split}
\label{eq:speeds}
    v_{\Sigma,0}&=a/\sqrt{L_J C_g} \\
    v_{\Delta,0}&=a/\sqrt{L_J\left( C_g + 2C_i \right)}
\end{split}
\end{align}
where $a$ is the unit cell length. We choose the line parameters such that $v_{\Sigma,0}/v_{\Delta,0}\simeq 3$, enabling the target processes in the band 4-10 GHz with a pump tuned from 2.5~GHz to 5~GHz. Due to the discrete nature of the unit cells, the wave velocity of each mode decreases when approaching a cut-off frequency above which no propagation is possible (see supplemental~\ref{sec:lumped-gaps}). Remaining parameters are chosen such that propagation modes exhibit a characteristic impedance of order 50~$\Omega$, allowing us to address them with minimum spurious reflections through  a tapered impedance transformer and a  hybrid coupler (see supplemental~\ref{sec:setup} and ~\ref{sec:layout} for details on the wiring setup and transformer geometry). \\

We may thus probe the $\Sigma$ mode transmission in presence of a continuous pump wave  applied from the right $\Delta$ port. In  Fig~\ref{fig2}, the magnitude of the probe transmission coefficient in each direction is reported in color against the pump frequency $\omega_{\textrm{pump}}$ and probe frequency $\omega_{\textrm{probe}}$, exhibiting gaps (blue lines) when a conversion process is phase-matched. Note that the coherently converted signal may be retrieved by detecting the reflected field at $\omega_{\textrm{probe}}\pm2\omega_{\textrm{pump}}$, which we verified for a few values  of $\omega_{\textrm{pump}}$  (see supplemental~\ref{sec:conversion}).\\

The circulation process, denoted by a marker Ci on the maps, corresponds to the leftmost feature in the left to right transmission (signal to idler upconversion in Fig.~\ref{fig0}b) and to the bottom  feature in right to left transmission (reverse idler to signal downconversion). Due to the discreteness of the device unit cells, these features curve at high frequency, as expected from the dispersion of the propagation modes. Moreover, a supplementary gap corresponding to the same circulation process opens at high frequency  through aliasing of a signal whose wavevector exceeds $\pi/a$ where $a$ is the length of a unit cell (top right feature labeled (Al) in right panel, with corresponding phase-matching diagram in inset). Crucially all these features are mostly visible in one direction of propagation only, demonstrating non-reciprocity. We attribute the opening of residual gaps in the opposite direction to spurious reflection of the pump within the device,  { unintentionally activating a conversion process in that direction. Quantitatively,  we estimate following an \textit{in situ} calibration of the device feedlines that the pump is reflected on the $\Delta$ propagation mode at the -2~dB level and leaks to the $\Sigma$ mode at the -4~dB level, which indicates that one of the junctions in the device is open (see supplemental~\ref{sec:model-open}).   Note that  the pump reflection could have been efficiently suppressed with a canceling pump applied in the reverse direction if the defective junction were  located close to the device ports. Unfortunately, time-domain reflectometry measurements indicate that the defective junction is located close to the middle of the line (see supplemental~\ref{sec:tdr}), making this strategy ineffective. As detailed in supplemental~\ref{sec:amplitudecalibration}, this situation effectively halves the  length of our active Josephson line and limits its performance.} \\

 {Another consequence of pump reflection is that forward and backward propagating pump waves may combine to activate the tunable coupling process.  The corresponding gap, denoted by a marker (Co) on the maps, is clearly visible at the center of both panels in Fig.~\ref{fig2}}, as expected from a reciprocal process.  All data pertaining to the tunable coupling process presented in section~\ref{sec:amp-dependency} was acquired by relying on pump reflection rather than by purposely applying a pump in both directions.  \\

The location of all gaps corresponding to the circulation and tunable coupling processes is accurately predicted (black dashed lines for expected gaps, gray dashed lines for residual gaps in the reverse transmission) by solving the phase-matching relations \ref{eq:energy} and \ref{eq:momentum} (see supplemental~\ref{sec:gap-shift}). In this model, we take   into account the discrete nature of the line unit cells and only leave as free fit parameters the wave velocities at low frequency---found to be within a few percent of their nominal design values---and the Josephson junctions plasma frequency  at 38.0~GHz.  {We further note the presence of a peak in the transmissions at $\omega_{\textrm{probe}}=\omega_{\textrm{pump}}$ and of a thin gap   lying close to the circulation process gap (visible left of the gray dashed line on the right panel). Both features are reproduced by numerical simulations based on a full discrete non-linear model~\cite{peng2022x} (see supplemental \ref{sec:extended-atten}). The former can be attributed to signal amplification arising from pump waves propagating in both directions, while the origin of the latter remains unclear at this stage.}

\section{\textbf{Device {performance} }}
\label{sec:amp-dependency}

\begin{figure}

\includegraphics[width=1\columnwidth]{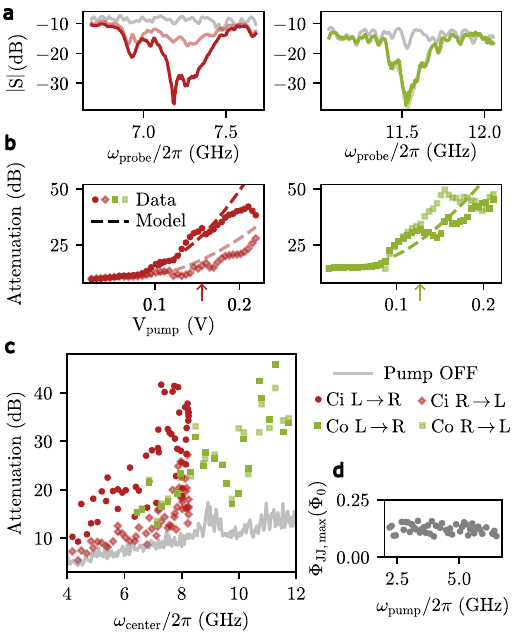}
    \caption{\label{fig3}{\bf Device  {performance. a,}}  {Forward and backward transmission (bright and dim colors, respectively) for a probe in the vicinity of the circulation gap (left panel, pump at $\omega_{\textrm{pump}}/2\pi=4.63$~GHz) or the tunable coupling gap (right panel, pump at $\omega_{\textrm{pump}}/2\pi=2.6$~GHz). For each process, the pump amplitude is adjusted at critical value (see text for details). The gray line shows transmission at zero pump amplitude. {\bf b,} Measured probe attenuation at the center gap frequency $\omega_{\textrm{center}}$, found by smoothing spectra as in (a) over a moving window of 40 MHz, against pump amplitude referenced at room temperature. Dashed lines represent predictions from a simple sideband model, the ratio between the room temperature pump amplitude and its value inside the device being independently calibrated (see supplemental \ref{sec:amplitudecalibration}). Attenuation at zero pump amplitude is attributed to dielectric loss, internal signal reflection by the defective junction and insertion loss of the hybrid couplers (see text for details).} {\bf c,} For each pump frequency, we locate the center frequencies of both processes. We adjust the pump amplitudes at  {critical value}, and report attenuations versus $\omega_{\textrm{center}}$. {\bf d,} estimated generalized flux across the device junctions, at the critical pump amplitudes, versus pump frequency (see supplemental~\ref{sec:phi-JJ}).
    }

\end{figure}
We further characterize our device  {performance} for non-reciprocal isolation and tunable coupling by recording the probe attenuation in both directions as a function of the applied pump amplitude. The attenuation is obtained, for a given pump amplitude and frequency, by smoothing the probe transmission spectrum {---referenced at the $\Sigma$ ports of the hybrid couplers---}over a 40 MHz window around the gap frequency and extracting its maximum depth  (see inset of Fig~\ref{fig3}a for a typical  spectrum before smoothing). In Fig~\ref{fig3}b, we report the attenuation measured for each process at a specific probe frequency and for a varying pump amplitude, referred by its voltage at room temperature. The low-frequency circulation process (labeled (Ci) in Fig~\ref{fig2} forward transmission) is activated by a pump at 4.63~GHz. It yields a desired  attenuation from the left to right port (dark red circles) and a residual  attenuation in the opposite direction resulting from pump reflection (light red diamonds). The tunable coupling process is activated by a pump at 2.6~GHz, and the corresponding attenuation is reported with green squares. Strikingly, all attenuations scale exponentially with the applied pump amplitude, in qualitative agreement  with the simple three-wave model presented in Fig~\ref{fig0}. \\

{Quantitatively,  for a line  of length $L$ in which waves propagate without reflection, this model predicts an attenuation  $2e^{-\alpha L}/(1+e^{-2\alpha L})$, where  the attenuation constant $\alpha$ depends on the considered process and scales quadratically with the traveling pump wave amplitude (see supplemental~\ref{sec:bands}). To capture the effect of wave scattering by the defective junction, we first calibrate the pump amplitude in each direction and on either side of the defect. We then solve the signal and idler wave dynamics in each section with joint boundary conditions set by the defect. We thus obtain the signal attenuation through the whole line (dashed lines in Fig~\ref{fig3}a, with a vertical offset adjusted to match the measured attenuation at zero pump amplitude), in good agreement with the measured attenuation for each activated process. For both processes, the measured attenuations start deviating significantly from  the prediction of our model  above a critical pump amplitude indicated by green and red arrows in Fig 4a. In particular, we measure a wideband drop in probe transmission for pump amplitudes above critical value, which is not captured by our model (see supplemental \ref{sec:extended-atten}).  Note that numerical simulations based on a full discrete non-linear model~\cite{peng2022x} do not yield significantly different results at the considered pump frequency (supplemental~\ref{sec:extended-atten}).  Over the  pump frequency range considered in this work, the critical pump amplitude corresponds  to an approximately constant peak amplitude $\Phi_{\textrm{JJ,max}}\sim0.12~\Phi_0$ for the generalized flux across the junctions  located right of the defective junction (see Fig.~\ref{fig3}d).}\\

 In Fig~\ref{fig3}c, we report the attenuations measured at the critical pump amplitude over a wide range of operating frequencies. Overall,  our device enables circulation with approximately 10~dB isolation---defined as the ratio of forward to backward attenuation---through the 5-8.5~GHz band, and approximately 20~dB in the 7-8.5~GHz band. Employed as a tunable coupler, it reaches an on/off  ratio  in transmission {from  10~dB to 20~dB in the 7-12~GHz band. The quoted values correspond to the attenuation at the center of each process gap. An attenuation reduced by a factor of two is maintained over a typical 200~MHz bandwidth around this frequency.} Note that the operating range in circulation is limited  by the modest electrical length of our device---resulting in  {performance} tailing off towards lower frequencies---and by the line cutoff frequencies---curving the dispersion relation towards higher frequencies. This range could thus be extended by increasing the number of cells in the device and by reducing the value of capacitances and inductances in each cell, which would also increase our device dynamic range, currently limited at  {-110~dBm} (see supplemental~\ref{sec:dyn-range}). Alternatively, one could invert the pump direction and leverage the other circulation processes (labeled Ci and Al in Fig~\ref{fig2}, backward transmission) to enable circulation at higher frequencies. As for the operating range in tunable coupling, it is only limited by the working bandwidth of various microwave components present in our setup (see supplemental~\ref{sec:setup}).\\

An important figure of merit for a circulator  is its insertion loss, corresponding to the right to left attenuation in Fig~\ref{fig3}c.  {When no pump is applied, we measure significant insertion loss (gray line) with a value of 8.5~dB at 7~GHz. Through independent calibration, we attribute 4.5~dB of this loss to the hybrid couplers and in-line filters and cables connecting to the device (see supplemental Fig~\ref{fig:setup}). We  estimate that signal reflection and leakage to the $\Delta$ mode---caused by the defective junction---contribute an additional 2~dB. The remaining 2~dB  are attributed to dielectric losses in the alumina, corresponding to a loss tangent of $\tan\delta\approx2\cdot10^{-3}$. Finally, when turning on the pump at critical value to activate the circulation process, we typically detect a few~dB   increase in insertion loss caused by reflected pump waves which activate the same target process  but in the reverse direction (light red diamonds). Note that all sources of internal loss could be significantly reduced through simple design changes---such as replacing parallel-plate capacitors with coplanar ones to eliminate lossy dielectric material ~\cite{wang2025high,chang2025josephson}---and improved fabrication to suppress internal reflections of pump and probe waves.}\\

\section{\textbf{Conclusion and outlook} }

We investigate a regime of traveling-wave parametric conversion by which the signal and idler waves propagate in opposite directions. The conversion process is phase-matched leveraging a slowly propagating pump wave and results in exponential attenuation of incoming signal waves as they travel down the device. The working frequency is tunable over a wide range by adjusting the pump frequency and conversion can be made either reciprocal or non-reciprocal  by applying the pump wave either from a single port or from both ports. Remarkably, our implementation based on a two-mode transmission line constructed from lumped capacitors and Josephson junctions performs close to state-of-the-art on-chip isolators~\cite{Chapman2017} in terms of bandwidth and isolation, despite the presence of a defect reflecting waves  in the middle of the line.\\

Looking forward, the device  {performance} could substantially improve through better-controlled fabrication, allowing to extend further the transmission line electrical length, to push its cutoff frequencies toward higher values and to increase its dynamic range while limiting spurious reflections. Alternative designs based on flux-pumped split-junctions~\cite{Zorin2019} or three-wave mixing elements~\cite{Zorin2016} will be explored to the same ends and to limit the participation of the dissipative dielectric material in the transmission mode carrying the signal. In terms of applications, robust on-chip isolators and tunable couplers compatible with standard fabrication techniques will benefit intermediate and large-scale superconducting circuit architectures. In particular, the circulation process we investigate in this work could be combined with standard traveling-wave amplification to achieve the long-standing goal of implementing a fully directional quantum limited amplifier~\cite{Liu2024}. Another exciting application  is the simulation  of condensed matter  in strong magnetic field~\cite{owens2022chiral,rosen2024implementing}. Multi-mode superconducting circuits endowed with non reciprocity   can thus exhibit non-trivial phases~\cite{qi2011topological,von202040}, and a two-mode non reciprocal circuit was recently proposed to implement a fully protected qubit~\cite{rymarz2021hardware}.

\section*{Acknowledgments}
The authors thank Alice \& Bob for providing the sample holder used in this experiment,  J. Palomo, M. Rosticher, A. Pierret and J-L. Smirr for assistance in fabrication, and F. Lecocq, M. Malnou and N. Roch for fruitful discussions. This work has been funded by the European Research Council (ERC) under the European Union’s Horizon 2020 research and innovation program (grant agreement No. 101042304),  by the French ANR-22-PETQ-0003 grant and by the ANR-22-PETQ-0006 grant under the ‘France 2030' plan.

\section*{Competing interests}
M.P. and P.C-I. are inventors of a patent related to the subject.

\section*{Data availability}
The experimental data and numerical simulations presented in this manuscript
are available from the corresponding authors upon request.

\section*{Author contributions}

M.P.  {and A.G.} fabricated the sample, performed the experiment and analyzed the data under the supervision of P.C-I.  {V. L. and T. B. performed complementary simulations and supporting calculations.} All authors contributed in designing the experiment and preparing the manuscript.

\section*{Correspondence}

Any correspondence should be addressed to philippe.campagne-ibarcq@inria.fr.

\bibliographystyle{naturemag}
\bibliography{library}


\clearpage
\onecolumngrid
\begin{center}
\textbf{\Large Supplemental Materials}
\end{center}

\setcounter{equation}{0}
\setcounter{section}{0}
\setcounter{figure}{0}
\setcounter{table}{0}
\setcounter{page}{1}
\renewcommand{\theequation}{S\arabic{equation}}
\renewcommand{\thefigure}{S\arabic{figure}}
\renewcommand{\thesection}{S\arabic{section}}

\section{Experimental setup}
 \label{sec:setup}

 \begin{figure}[h]
     \centering
     \includegraphics[width=1\linewidth]{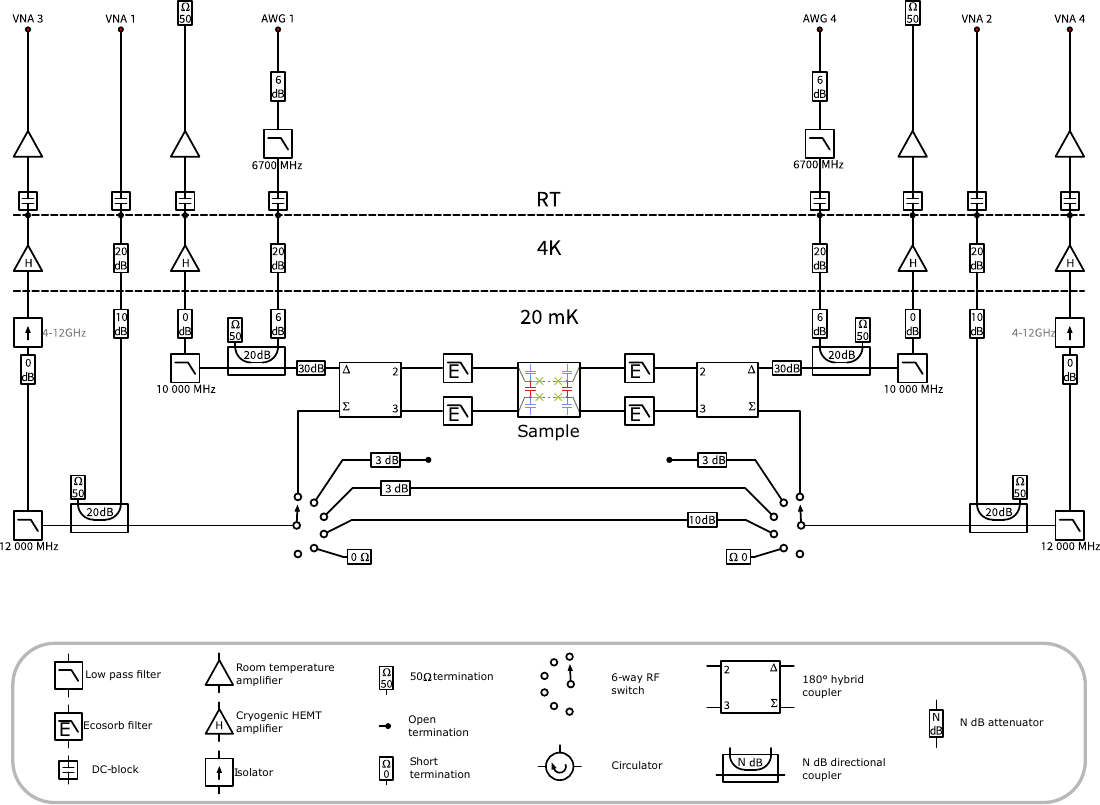}
     \caption{{\bf Measurement setup.} See text for details.}
     \label{fig:setup}
 \end{figure}
 
 The sample was measured using a vector network analyzer (VNA, model P5024A from Keysight) in a dilution refrigerator with a base temperature of 10mK. An arbitrary waveform generator (AWG, model M8195A from Keysight, with a sampling rate of 65GSa/s) was used to generate the pump waves, ensuring a spectrally pure wave, with controlled relative phase and harmonics at a level better than -20dBc.\\
 
 The $\Sigma$ and $\Delta$ eigenmodes of the sample are addressed from both sides through 180° hybrid couplers and directional couplers. In practice, the pump is sent through the $\Delta$ port of the rightmost hybrid coupler or through the $\Delta$ port of each hybrid coupler (depending on circulator or coupler configuration), and the probe tone from the VNA is sent to the $\Sigma$ port of one or the other hybrid coupler. The sample is held inside layers of mu-metal, aluminum and copper shielding.  Each cable connecting to the system is filtered against infrared radiations. The output lines of the $\Sigma$ ports are routed through isolators (model LNF-ISIS-C412A from Low Noise Factory) and commercial HEMT amplifiers (model LNF-LNC0.3\_14B from Low Noise Factory). The output lines of the $\Delta$ ports are routed through a 30~dB attenuator instead of an isolator to protect against the amplifier thermal noise, as circulators covering the pump frequency range were not available.\\
 
Finally, microwave switches on either side of the $\Sigma$ measurement lines allow to measure calibration standards instead of the sample. The minimal set of reflect, thru and attenuated line standards are present to effectively shift the VNA reference plane to the input of the hybrid couplers $\Sigma$ ports~\cite{EulTMR, Engen1979}. It is noteworthy that the post-calibration reference planes are still far from the two-mode Josephson line at the core of the sample: it is a limitation of using a four-port sample addressed through hybrid couplers instead of a two-port sample. Undesirable losses from the filters, hybrid couplers, and long cabling (only 1m cables were available) are still present. However, this calibration is what allows precise knowledge of the attenuation and phase of the lines, and in turn accurate time domain reflectometry, see appendix~\ref{sec:tdr}.\\

Spectrum analyzer measurements, were performed by using one of the two VNA ports as a continuous wave source to send the input  tone while the analyzer (model SA124B from Signal Hound) is plugged at an output in place of the other VNA port. 

\section{Device layout, impedance matching and equations of motion }

\subsection{Impedance matching}
\label{sec:layout}

\begin{figure}[h]
    \centering
    \includegraphics{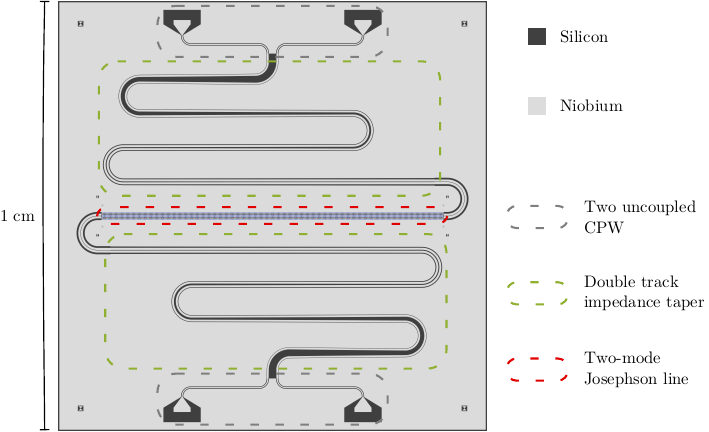}
    \caption{{\bf Device layout.} The device features two launchpads at the top of the chip, and two at the bottom. Starting from two launchpads, two standard CPW lines merge together, at the start of the Klopfenstein impedance taper. The lines are slowly coupled together in the taper, which is meandered for increased length. The taper is then connected to the nonlinear transmission line, in the middle of the sample.}
    \label{fig:chip}
\end{figure}

In this section, we describe the device layout, which is constrained by impedance considerations. The device must be connected to the standard 50~$\Omega$ coaxial lines of the cryostat. However, given the electrical layout of Fig~\ref{fig:circuit-theory}, the lines are capacitively coupled, and it is impossible to have both propagating mode impedances at 50$\Omega$ at the same time unless additional inductive coupling is used~\cite{Naghiloo2021}, which was not attempted. More precisely, defining the eigenmode voltages and currents as $V_{\Delta}=(V_2-V_1)/2$, $V_{\Sigma}=(V_2+V_1)/2$, $i_{\Delta}=(i_2-i_1)/2$, $i_{\Sigma}=(i_2+i_1)/2$, the mode impedances are found to be~\cite{ClaytonBook}:

\begin{align}
    Z_{\Sigma}&=\sqrt{\frac{L_J}{C_g}}\\
    Z_{\Delta}&=\sqrt{\frac{L_J}{C_g+2C_i}}
\end{align}

Note that with this $\Sigma$/$\Delta$ eigenmode convention, two identical, single-mode, uncoupled, 50~$\Omega$ lines will, when considered together, verify $Z_{\Sigma}=Z_{\Delta}=50~\Omega$.\\

\begin{figure}[h]
    \centering
    \includegraphics{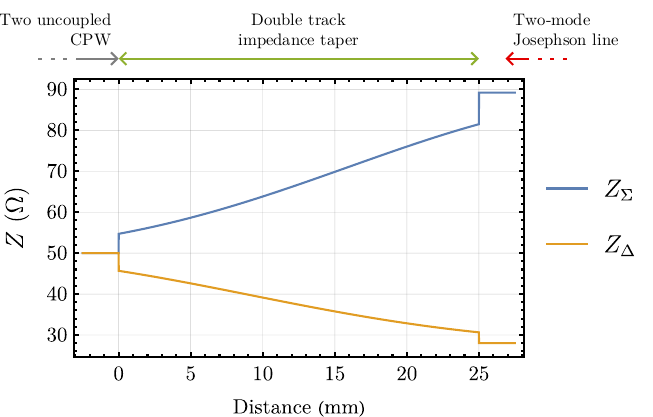}
    \caption{{\bf Impedance profile of the device Klopfenstein taper} as a function of distance to uncoupled CPW lines. Below 0mm distance, the multiconductor line is uncoupled, and composed of two CPW lines of impedance 50~$\Omega$ each. They are suddenly lightly coupled (at distance 0~mm), then the coupling is increased with position, until the point (at 25~mm) where the coupling is abruptly adjusted close to the final Josephson transmission line characteristic impedances.}
    \label{fig:impedance}
\end{figure}

In our device, we made the choice of $L_J$, $C_g$ and $C_i$ values such that $\sqrt{Z_{\Sigma}Z_{\Delta}} \simeq 50~\Omega$. One mode has an impedance slightly above 50~$\Omega$ ($Z_{\Sigma}=85~\Omega$) and the other slightly below ($Z_{\Delta}=27~\Omega$). We then match both modes to 50~$\Omega$ simultaneously using a Klopfenstein impedance taper~\cite{Pozar}. To achieve this, we slowly vary multiple dimensions of the Ground-Signal-Signal-Ground coplanar waveguides that connect the non-linear transmission line to the launchpads along the length of the taper, see Fig~\ref{fig:chip}. Note that there is an impedance jump caused by the sudden introduction of a middle ground plane, where the multiconductor coupled line parts into two uncoupled coplanar waveguide lines. There is another jump at the other extremity of the taper. This is desired in the specific case of the Klopfenstein taper, see the profile in Fig~\ref{fig:impedance}. Finally, we note that the final taper impedances  ($\tilde{Z}_{\Sigma}=89~\Omega$ and $\tilde{Z}_{\Delta}=28~\Omega$) are slightly higher than the mode impedances. This proceeds from an error in the final parameters adjustment during fabrication. The resulting wave reflection at the sample  ports are negligible (below -30~dB).

\subsection{Lagrangian for the two-mode Josephson line}
\label{sec:lagrangian}

\begin{figure}[h]
    \centering
\includegraphics[width=0.5\columnwidth]{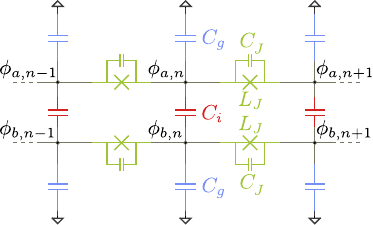}
    \caption{{\bf Two elementary cells of the two-mode Josephson line.} The fluxes $\phi_{a,i}$ and $\phi_{b,i}$ are defined at the nodes of the corresponding cell $i$.}
    \label{fig:circuit-theory}
\end{figure}

In this section, we derive the equations of motion for signals traveling down the two-mode Josephson line. The line's elementary cell is depicted in Fig.~\ref{fig:circuit-theory}. Neglecting the contribution of the first and last cell of the line, the Lagrangian of the line reads
\begin{equation}
\begin{aligned}
\mathcal{L}=&\sum_n E_J \cos(\frac{\phi_{a,n+1}-\phi_{a,n}}{\varphi_0}) + E_J \cos(\frac{\phi_{b,n+1}-\phi_{b,n}}{\varphi_0})\\
&\qquad+ \frac{C_g}{2} (\dot{\phi}_{a,n}^2+\dot{\phi}_{b,n}^2)+\frac{C_i}{2} (\dot{\phi}_{a,n}-\dot{\phi}_{b,n})^2+\frac{C_J}{2} \big((\dot{\phi}_{a,n+1}-\dot{\phi}_{a,n})^2 + (\dot{\phi}_{b,n+1}-\dot{\phi}_{b,n})^2 \big)
\end{aligned}
\end{equation}
 We let $\varphi_{a,n}=\phi_{a,n}/\varphi_0$, $\varphi_{b,n}=\phi_{b,n}/\varphi_0$ (where $\varphi_0=\hbar/(2e)$ is the reduced flux quantum), $L_J=\varphi_0^2/E_J$, $\omega_g=1/\sqrt{L_JC_g}$ and $\mu=1+2C_i/C_g$ and express this Lagrangian as a function of the symmetric and antisymmetric coordinates $\varphi_{\Sigma,n}=(\varphi_{a,n}+\varphi_{b,n})/2$ and $\varphi_{\Delta,n}=(\varphi_{a,n}-\varphi_{b,n})/2$ as
\begin{equation}
\mathcal{L}=\frac{\varphi_0^2}{L_J} \sum_n 2\cos(\varphi_{\Sigma,n+1}-\varphi_{\Sigma,n})\cos(\varphi_{\Delta,n+1}-\varphi_{\Delta,n}) + \frac{1}{\omega_g^2} \dot{\varphi}_{\Sigma,n}^2 +\frac{\mu}{\omega_g^2} \dot{\varphi}_{\Delta,n}^2+\frac{1}{\omega_J^2} \big((\dot{\varphi}_{\Sigma,n+1}-\dot{\varphi}_{\Sigma,n})^2 + (\dot{\varphi}_{\Delta,n+1}-\dot{\varphi}_{\Delta,n})^2 \big)
\label{eq:Lagragian-sym}
\end{equation}
The Euler-Lagrange equations then read
\begin{equation}
\begin{aligned}
&\frac{1}{\omega_g^2} \ddot{\varphi}_{\Sigma,n}+\frac{1}{\omega_J^2} (2\ddot{\varphi}_{\Sigma,n}-\ddot{\varphi}_{\Sigma,n+1}-\ddot{\varphi}_{\Sigma,n-1})\\
 & \qquad \qquad=\sin(\varphi_{\Sigma,n+1}-\varphi_{\Sigma,n})\cos(\varphi_{\Delta,n+1}-\varphi_{\Delta,n})-\sin(\varphi_{\Sigma,n}-\varphi_{\Sigma,n-1})\cos(\varphi_{\Delta,n}-\varphi_{\Delta,n-1})\\
&\frac{\mu}{\omega_g^2} \ddot{\varphi}_{\Delta,n}+\frac{1}{\omega_J^2} (2\ddot{\varphi}_{\Delta,n}-\ddot{\varphi}_{\Delta,n+1}-\ddot{\varphi}_{\Delta,n-1})\\
 & \qquad \qquad=\sin(\varphi_{\Delta,n+1}-\varphi_{\Delta,n})\cos(\varphi_{\Sigma,n+1}-\varphi_{\Sigma,n})-\sin(\varphi_{\Delta,n}-\varphi_{\Delta,n-1})\cos(\varphi_{\Sigma,n}-\varphi_{\Sigma,n-1})
\end{aligned}
\label{eq:eomlumped}
\end{equation}

\section{Sideband model in the continuous approximation}
\label{sec:bands}

In this section, we derive simplified analytical formula for the attenuation induced by the circulation and tunable coupling processes, inspired by Ref.~\cite{Peng2022}. These formulas  are valid in a limited frequency and amplitude range for the pump and signals. In detail, the hypotheses that we make are
\begin{enumerate}
\item The length $a$ of the unit cell is much shorter than all wavelengths involved. This is equivalent to saying that the signal and idler frequencies  are much smaller than the $\Sigma$ mode cutoff frequency and that the pump frequency is much smaller than the $\Delta$ mode cutoff frequency (see Sec.~\ref{sec:lumped-gaps}). 
\item The device electrical length is longer than all wavelengths involved. This allows us to consider only phase-matched processes as relevant.
\item The reduced wave amplitudes are much smaller than 1. This is equivalent to saying that the phase-drop across each junction is much smaller than the quantum of flux. 
\item There are no internal reflections of the waves except for the expected reflection and frequency conversion of the signal induced by the processes.
\end{enumerate}

\subsection{Continuous equations of motion}

Given the hypothesis (1), we consider the continuous limit of the equations of motion \eqref{eq:eomlumped}
\begin{equation}
\begin{aligned}
\frac{1}{\omega_g^2} \ddot{\varphi}_{\Sigma}&=a\Big( \sin(a\varphi_{\Sigma}') \cos(a\varphi_{\Delta}') \Big)'\\
\frac{\mu}{\omega_g^2} \ddot{\varphi}_{\Delta}&=a\Big( \sin(a\varphi_{\Delta}') \cos(a\varphi_{\Sigma}') \Big)'
\end{aligned}
\label{eq:eom}
\end{equation}
where we used the convention that $\dot{\_}$ designates a partial derivative with respect to time and $\_'$ a partial derivative with respect to position.\\

When the line is driven along $\varphi_{\Delta}$ with a  pump at $\omega_P$ with amplitude $\epsilon_P \ll 1$ and along $\varphi_{\Sigma}$ with a weak signal at $\omega_S$ of amplitude $\epsilon_S \ll \epsilon_P$ (hypothesis (3)), 
we have $a\varphi'_{\Sigma}\ll a\varphi'_{\Delta}$, as well as $a\varphi''_{\Sigma}\ll a\varphi''_{\Delta}$. Again, assuming the length a is much shorter than all wavelengths involved we expand to second order in $a\varphi'_{\Delta}$ and $a^2\varphi''_{\Delta}$ and to first order in $a\varphi'_{\Sigma}$ and $a^2\varphi'_{\Sigma}$ in Eq.~\eqref{eq:eom}. We thus get 
\begin{equation}
\begin{aligned}
\frac{1}{\omega_g^2} \ddot{\varphi}_{\Sigma}&=a^2 \varphi_{\Sigma}'' - \frac{a^4}{2}\varphi_{\Sigma}'' \varphi_{\Delta}'^2-a^4 \varphi_{\Sigma}' \varphi_{\Delta}' \varphi_{\Delta}'' \\
\frac{\mu}{\omega_g^2} \ddot{\varphi}_{\Delta}&=a^2 \varphi_{\Delta}''
\end{aligned}
\label{eq:eomsimple}
\end{equation}
The second equation becomes linear so that the pump propagating along $\varphi_{\Delta}$ is stiff  and reads 
\begin{equation}
\label{eq:pump-ansatz}
\varphi_{\Delta}(x,t)=\epsilon_P e^{i(\omega_Pt -k_Px )} + c.c 
\end{equation}
where $k_P=-\omega_P/v_{\Delta}$ and  $v_{\Delta}=a \omega_g/\sqrt{\mu} $ for a backward propagating pump (here we consider the ideal case where the pump is applied from the right sample port and that there is no internal reflections following hypothesis (4)). Thus, by truncating the Taylor series of $a\varphi'_{\Sigma}$, $a^2\varphi''_{\Sigma}$, $a\varphi'_{\Delta}$ and $a^2\varphi''_{\Delta}$, we  have neglected auto-modulation of the pump and cross-modulation from the signal (see Sec.~\ref{sec:nonlinear-gaps} for corrections induced by phase modulation of the signals by the pump). This solution can then be injected in the first equation in \eqref{eq:eomsimple} where it modulates the properties of the  $\Sigma$ mode via four-wave mixing processes involving two pump photons. \\

Following the general method laid out in \cite{Yaakobi2013}, we  look for solutions of the equation of motion for $\varphi_{\Sigma}$ under the form of a combination of sideband signals at $\omega_S + 2k \omega_P$ with $k$ integer. When $\omega_S$ is close to $\omega_P(\frac{v_{\Sigma}}{v_{\Delta}}-1)$ as prescribed in Eq.~\eqref{eq:signal-freq} the dominant sideband is  at $\omega_I=\omega_S+2\omega_P$ ($k=-1$), and we refer to it as the idler---all other processes, and in particular the one yielding a signal at $\omega_I=\omega_S-2\omega_P$ are not phase-matched in our approach. Note that phase-matching is only relevant under hypothesis (2). We now focus on a simple model in which all other sideband signals are neglected. We thus look for a solution of the equation of motion under the form
 \begin{equation}
 \label{eq:phi-sigma-form}
 \varphi_{\Sigma}(x,t)=\epsilon_S(x) e^{i(\omega_S t - k_S x) } + \epsilon_I(x) e^{i (\omega_I t - k_I x)} + c.c.
 \end{equation}
 where $k_S=\omega_S/v_{\Sigma}$ and  $k_I=-\omega_I/v_{\Sigma}$ with $v_{\Sigma}=a \omega_g $ for a forward propagating signal and backward propagating idler, and $\epsilon_S(x)$ and $\epsilon_I(x)$ are slowly varying functions ($\epsilon'_{S,I}\ll |k_{S,I}| \epsilon_{S,I}$) that we want to solve for.\\

\subsection{Circulation process}
\subsubsection{In-band attenuation}

\begin{figure}[h]
    \centering
    \includegraphics{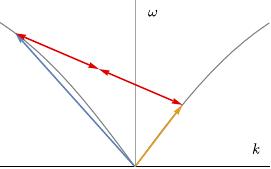}
    \caption{{\bf Phase-matching diagram of the circulation process.} A forward propagating signal (orange) can be converted to a backward propagating idler (blue) through the interaction with two backward propagating pump photons (red).}
    \label{fig:isol-supmat}
\end{figure}

We first consider the case of a signal applied at the center frequency of the circulator working bandwidth. As a reminder, in this process a forward propagating signal can be converted to a backward propagating idler through the interaction with two backward propagating pump photons, see Fig.~\ref{fig:isol-supmat}.
The center frequency of the circulator is determined from the phase-matching relations $\omega_S=\omega_I-2\omega_P$ and $k_S+2k_P=k_I$, yielding $\omega_S=\omega_P(\frac{v_{\Sigma}}{v_{\Delta}}-1)$. Letting $\theta_Z=\omega_Z t-k_Z x$ for $Z=S,~I,~P$, we recast the first line in the equation of motion \eqref{eq:eomsimple} as
\begin{equation}
\begin{aligned}
0=\sum_{Z=S,I} \Big(\quad &(\frac{\omega_Z^2}{\omega_g^2}-a^2k_Z^2) &\epsilon_Z e^{i\theta_Z}&  \\
 +&\frac{a^4}{2} ( k_Z^2  \varphi_{\Delta}'^2  +2ik_Z\varphi_{\Delta}'\varphi_{\Delta}'') &\epsilon_Z e^{i\theta_Z} &\\
+&  a^2 (-2ik_Z +a^2 i k_Z \varphi_{\Delta}'^2-a^2 \varphi_{\Delta}'\varphi_{\Delta}'' ) & \epsilon_Z' e^{i\theta_Z}&\\
+& a^2 (1-\frac{a^2}{2}\varphi_{\Delta}'^2)& \epsilon_Z'' e^{i\theta_Z}\\
&&& \quad+c.c. \Big)
\end{aligned}
\end{equation}
In this expression, 
\begin{itemize}
\item The terms of the first line sum up to zero. 
\item Using that $| \epsilon_Z''|\ll |k_Z  \epsilon_Z'|$ (slowly varying function), we can neglect terms on the fourth line compared to terms on the third line. 
\item Using that $a^2 |\varphi_{\Delta}'^2| = a^2|k_P^2 \epsilon_P^2 |\ll 1$ and $a^2 |\varphi_{\Delta}' \varphi_{\Delta}''|=a^2|k_P^3 \epsilon_P^2 |\sim a^2|k_P^2 k_Z \epsilon_P^2 |\ll k_Z$ , we can neglect the center and rightmost terms in the third line.
\end{itemize}
Altogether, we are left with 
\begin{equation}
\sum_{Z=S,I}  2ik_Z  \epsilon_Z' e^{i\theta_Z} +c.c. = \sum_{Z=S,I} \frac{a^2}{2} ( k_Z^2  \varphi_{\Delta}'^2  +2ik_Z\varphi_{\Delta}'\varphi_{\Delta}'') \epsilon_Z e^{i\theta_Z} +c.c.
\label{eq:eomsimple2}
\end{equation}
Expressing $\varphi_{\Delta}'=-i k_P\epsilon_P e^{i\theta_P} + c.c$ and $\varphi_{\Delta}''=- k_P^2\epsilon_P e^{i\theta_P}+c.c$ and given that $\theta_S+2\theta_P=\theta_I$, we can then identify resonant terms of the form $A e^{i\theta_S}=B e^{i\theta_I - 2\theta_P}$ and $C e^{i\theta_I}=D e^{i\theta_S + 2\theta_P}$ (and their complex conjugate) as contributing dominantly to the dynamics. We thus get two coupled differential equations for $\epsilon_S$ and $\epsilon_I$:
\begin{equation}
\begin{aligned}
 \label{eq:eomprefinal}
2i k_S \epsilon_S'&=\frac{a^2}{2}~(-k_I^2 k_P^2 + 2 k_I k_P^3)~\epsilon_P^{\ast 2} \epsilon_I \\
2i k_I \epsilon_I'&=\frac{a^2}{2}~(-k_S^2 k_P^2 - 2 k_S k_P^3)~\epsilon_P^2 \epsilon_S
\end{aligned}
\end{equation}
and this system boils down to 
\begin{equation}
\begin{aligned}
 \epsilon_S'&=i\frac{a^2}{4}~\epsilon_P^{\ast 2} k_P^2 k_I \epsilon_I \\
 \epsilon_I'&=i\frac{a^2}{4}~ \epsilon_P^2 k_P^2 k_S \epsilon_S
 \label{eq:eomfinal}
\end{aligned}
\end{equation}
One may trivially decouple this system into two separated second order differential equations. Given that $k_I k_S<0$ in our approach, the solutions are combinations of increasing and decreasing exponential functions. In the infinite line case, since the total number of excitations in the idler and signal is preserved by the conversion process, increasing exponentials are unphysical and the solutions read $ \epsilon_S(x)=\epsilon_S(0)e^{-\alpha x}$ and $ \epsilon_I(x)=\epsilon_I(0)e^{-\alpha x}$ where the attenuation constant is $\alpha=\frac{a^2}{4}k_P^2 \sqrt{-k_I k_S}|\epsilon_P|^2$. Alternatively, we can solve analytically these equations with the boundary conditions $\epsilon_S(0)=\epsilon_{S,0}$ and $\epsilon_I(L)=0$, where $L=N_{cell}a$ is the length of the device. We find:
\begin{align}
    \epsilon_S(x)&=\epsilon_{S,0}
    \frac{e^{-\alpha x}+e^{-\alpha (2L-x)}}
    {1+e^{-2\alpha L}}\\
    \epsilon_I(x)&=-i e^{2i\mathrm{Arg}(\epsilon_P)}\epsilon_{S,0}\sqrt{\frac{k_S}{-k_I}}
    \frac{e^{-\alpha x}-e^{-\alpha (2L-x)}}
    {1+e^{-2\alpha L}}
\end{align}

In particular, at the end of the device, we find the total attenuation:
\begin{equation}
\label{eq:final-isol}
    \frac{\epsilon_S(L)}{\epsilon_{S,0}}=\frac{2 e^{-\alpha L}}{1+e^{-2\alpha L}}
\end{equation}

In Sec.~\ref{sec:extended-atten} we compare the predictions from this analytical formula to those from a discrete non-linear model for frequencies approaching the modes cutoff frequencies.

\subsubsection{Working bandwidth in the continuous model}

To find the isolation bandwidth of our device, we consider the case where the signal frequency is slightly detuned from the optimal working point $\omega_S=\omega_P(\frac{v_{\Sigma}}{v_{\Delta}}-1)+\delta$ with $\delta \ll \omega_{S,I,P} $. We still consider an idler band at $\omega_I=\omega_S+2\omega_P$ (enforced by energy conservation), but the conversion process is imperfectly phase-matched with $k_S+2k_P-k_I=\kappa$ with $\kappa=2\delta/v_{\Sigma}$. We can still identify quasi-resonant processes from Eq.~\eqref{eq:eomsimple2} and find the coupled differential equations for $\epsilon_S$ and $\epsilon_I$ to be
\begin{equation}
\begin{aligned}
\label{eq:detuned}
 \epsilon_S'&=i\frac{a^2}{4}~e^{-i\kappa x}~ \epsilon_P^{\ast 2} k_P^2 k_I \epsilon_I \\
 \epsilon_I'&=i\frac{a^2}{4}~e^{i\kappa x} \epsilon_P^2 k_P^2 k_S \epsilon_S 
\end{aligned}
\end{equation}
This system may be decoupled as  second order equations, the  equation for $\epsilon_S$ reading
\begin{equation}
\epsilon_S''-i \kappa \epsilon_S' - \alpha^2\epsilon_S=0
\end{equation}
This equation admits propagating  solutions  if $\kappa^2-4\alpha^2 >0$. We thus find the isolation bandwidth limit  to be set by $\kappa=2\alpha$ and we estimate a total bandwidth $B=\frac{a^2}{2}k_P^2 |\epsilon_P|^2\sqrt{\omega_I\omega_S}$. Alternatively, the coupled equations~\eqref{eq:detuned} may be solved analytically, but the resulting expression is cumbersome. In Sec.~\ref{sec:exp-bandwidth} we compare predictions from this continuous, analytical model---enriched to account for pump reflection by the defective junction---with numerical simulations based on a more realistic model accounting for the discrete structure of the line and the full non-linearity of its unit cells.

\subsection{Tunable coupling process}
\label{sec:coupler-sup}
\begin{figure}[h]
    \centering
    \includegraphics{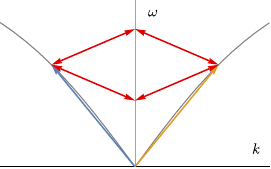}
    \caption{{\bf Phase matching paths of the tunable coupling process.} The signal (orange) and idler (blue) arrows can be arranged in a closed form through the top two pump arrows (red), or the bottom two. This is equivalent to the two possible orders of terms of $\varphi_{\Delta}$ in~\eqref{eq:eomsimple2}.}
    \label{fig:coupler-2paths}
\end{figure}

We proceed in a similar fashion for a signal applied at the center frequency of the coupler working bandwidth. As a reminder, in this reciprocal process, a photon in the $\Sigma$ mode interacts with a photon of either the backward or forward propagating pump, leading to the reflection of both photons, see Fig.~\ref{fig:coupler-2paths}. Considering a forward propagating photon in the $\Sigma$ mode, close to $\omega_S=\omega_P\frac{v_{\Sigma}}{v_{\Delta}}$, the dominant sideband is a counterpropagating wave at $\omega_I=\omega_S$. We still use $\varphi_{\Sigma}$ form of eq~\eqref{eq:phi-sigma-form}, with the new values for the idler wave. We now have to consider two counter-propagating pumps:
\begin{equation}
    \varphi_{\Delta}(x,t)=\epsilon_{P,\textrm{fw}} e^{i(\omega_Pt -k_{P,\textrm{fw}}x )} + \epsilon_{P,\textrm{bw}} e^{i(\omega_Pt -k_{P,\textrm{bw}}x )} + c.c 
\end{equation}
where $k_{P,\textrm{fw}}=\omega_P/v_{\Delta}=-k_P$ and $k_{P,\textrm{bw}}=-\omega_P/v_{\Delta}=k_P$.
In that case, we have $k_S+k_{P,\textrm{bw}}=k_I+k_{P,\textrm{fw}}$. Eq~\eqref{eq:eomsimple2} is still valid, and after identifying the dominant resonant terms, we get two coupled differential equations:
\begin{equation}
\begin{aligned}
2i k_S \epsilon_S'&=\frac{a^2}{2}~(-2k_I^2 k_P^2 + 4 k_I k_P^3)~\epsilon_{P,\textrm{fw}}\epsilon_{P,\textrm{bw}}^{\ast} \epsilon_I \\
2i k_I \epsilon_I'&=\frac{a^2}{2}~(-2 k_S^2 k_P^2 - 4 k_S k_P^3)~\epsilon_{P,\textrm{bw}}\epsilon_{P,\textrm{fw}}^{\ast} \epsilon_S
\end{aligned}
\end{equation}
which are very similar to coupled equations~\eqref{eq:eomprefinal}, up to a change $\epsilon_{P}^2\rightarrow 2 \epsilon_{P,\textrm{bw}}\epsilon_{P,\textrm{fw}}^{\ast}$. The factor 2 stems from the symmetry of the product terms $\varphi_{\Delta}$ in~\eqref{eq:eomsimple2} and can be understood as an alternative path in the phase matching diagram, as shown in Fig~\ref{fig:coupler-2paths}.

The coupled equations are solved in the same manner as the circulating process case.

\section{Lumped and non-linear  corrections to the dispersion relations }

Our device is made of discrete, non-linear lumped elements. When the wavelength of the traveling wave is much larger than the length of a cell  (hypothesis (1) in the introduction of Sec.~\ref{sec:bands}) and all waves have a reduced amplitude much smaller than 1 (hypothesis (3)) then the velocity of a wave does not depend on its frequency and amplitude, and is given by  Eq~\eqref{eq:speeds}. If either of these assumptions breaks down, the wave velocity is no longer constant.\\

In Sec.~\ref{sec:lumped-gaps} we derive the modified dispersion relations when hypothesis (1) is not satisfied. The wave velocity renormalization is then a purely linear effect stemming from the discrete structure of the line. In Sec.~\ref{sec:nonlinear-gaps},  we derive the modified dispersion relations when hypothesis (3) is not satisfied. The wave velocity renormalization then comes from phase modulation by the pump.

\subsection{Lumped element corrections}
\label{sec:lumped-gaps}

We start from the discrete equations of motion derived in Sec.~\ref{sec:lagrangian} and first consider a wave of frequency $\omega$ and wavevector $k$ traveling on the $\Sigma$ mode of the line. We assume that the wave amplitude $\epsilon$ to be small and constant during its propagation (no active conversion or amplification process). Injecting  $\varphi_{\Sigma,n}(t)=\epsilon e^{i \theta_n} +c.c.$ (with $\theta_n=\omega t -k na$, $a$ is the length of a unit cell) and $\varphi_{\Delta,n}(t)=0$ in the first line of Eq.~\eqref{eq:eomlumped}, truncating the cosine and sine functions at first order and equating left and right-hand side terms whose phase rotate as $\theta_n$ we get
\begin{equation}
\frac{\omega^2}{2\omega_g^2}+\frac{\omega^2}{\omega_J^2}(1-\cos(ka))=1-\cos(ka)
\label{eq:fordisp}
\end{equation}
 We can write a similar equation for a low-amplitude wave propagating on the $\Delta$ mode. From these, we get the modified dispersion relation (also derived in Ref.~\cite{Greco2013})
\begin{align}
\begin{split}
    k_{\Sigma}(\omega)&=\frac{1}{a}\arccos\left( 1+\frac{C_g L_{J}\,\omega^2}{-2+2C_J L_{J}\, \omega^2} \right)\\
    k_{\Delta}(\omega)&=\frac{1}{a}\arccos\left( 1+\frac{(C_g+2C_i) L_{J}\omega^2}{-2+2C_J L_{J}\, \omega^2} \right)
\end{split}
\label{eq:correc-lumped}
\end{align}

\begin{figure}[h]
    \centering
    \includegraphics{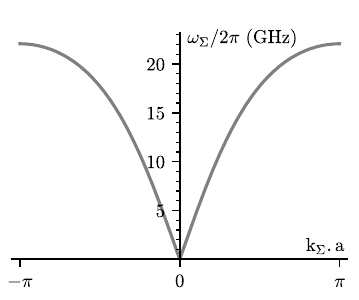}
    \caption{{\bf Dispersion relation of the sample $\Sigma$ mode.}  At low frequencies, the simplified formula for phase velocity $\omega/k$ of equation ~\eqref{eq:speeds} applies. When the frequency becomes comparable with the cut-off of the line, the phase velocity diminishes following Eq.~\eqref{eq:correc-lumped}. The cut-off frequency corresponds to a phase $\pm\pi$ across a cell, or equivalently half a wavelength.}
    \label{fig:disp-rel}
\end{figure}
The cut-off frequencies are given by:
\begin{equation}
\begin{aligned}
    \omega_{\Sigma,co}&=\frac{2}{\sqrt{L_{J}(C_g+4C_J)}}\\
    \omega_{\Delta,co}&=\frac{2}{\sqrt{L_{J}(C_g+2C_i+4C_J)}}
\end{aligned}
\end{equation}
The dispersion relation for the $\Sigma$ mode, computed using the fitted value for the circuit parameters, is plotted in Fig~\ref{fig:disp-rel}. With these parameters, the cut-off frequencies are $\omega_{\Sigma,co}/(2\pi)=22.56$~GHz and $\omega_{\Delta,co}/(2\pi)=9.44~$GHz.

\subsection{Non-linear corrections}
\label{sec:nonlinear-gaps}

\subsubsection{Self-modulation}
We now consider the same wave traveling on the $\Sigma$ mode but with a non negligible reduced amplitude $\epsilon$. We look for leading order corrections to its propagation velocity due to the line non-linearity, known as self phase-modulation~\cite{OBrien2014}. We still assume that no phase-matched process is active, such that the wave amplitude is constant throughout the line.\\

We first re-write the arguments of the sine functions in Eq.~\eqref{eq:eomlumped} as
\begin{equation}
\varphi_{\Sigma,n+1}-\varphi_{\Sigma,n}=4\epsilon \sin(\frac{k a}{2})\sin(\theta_n-\frac{ka}{2})
\label{eq:phasedrop0}
\end{equation}
where $\theta_n=\omega t-k an$. We then, as  in the linear case of the previous section, equate  terms with phase rotating as $\theta_n$ from left and right-hand side in the equation of motion \eqref{eq:eomlumped}. However, instead of truncating the sine function to first order, we here use the order 1 term of its Jacobi-Anger expansion. This amounts to substituting
\begin{equation}
\frac{1}{L_J}4\epsilon \sin(\frac{ka}{2}) \rightarrow \frac{1}{L_J} 2 J_1(4\epsilon \sin(\frac{ka}{2}))
\label{eq:SPM}
\end{equation}
where $J_n$ is the $n$-th Bessel function of the first kind. Therefore,  self phase-modulation of a wave is equivalent to a renormalization of the Josephson inductance. To leading order in $\epsilon$, it reads $L_{J,SPM}(\epsilon)=L_J(1+2\sin^2(\frac{ka}{2})\epsilon^2)$. We find the same renormalization for a wave propagating on the $\Delta$ mode.

\subsubsection{Cross-modulation}

We now consider the renormalization of the velocity of a weak amplitude wave (with frequency $\omega_2$, wavevector $k_2$ and amplitude $\epsilon_2$) by a large amplitude wave (with frequency $\omega_1 $, wavevector $k_1 $ and amplitude $\epsilon_1 \gg \epsilon_2$).\\

We first assume that the wave 2 propagates on the $\Delta$ mode, the wave 1 on the $\Sigma$ mode. Starting from the first line of Eq.~\eqref{eq:eomlumped}, we inject $\varphi_{\Delta,n}=\epsilon_1 e^{i\theta_{1,n}}+c.c.$ and $\varphi_{\Sigma,n}=\epsilon_2 e^{i\theta_{2,n}}+c.c.$ (still with the convention $\theta_{i,n}=\omega_i t - k_i an$). When equating terms rotating as $\theta_{1,n}$ on both sides of the equation and assuming that $\omega_1 \neq \omega_2 $  we find the velocity renormalization due to cross phase-modulation to be equivalent to substituting
\begin{equation}
\frac{1}{L_J} \rightarrow \frac{1}{L_J}  J_0(4\epsilon \sin(\frac{ka}{2}))
\label{eq:XPM}
\end{equation}
To leading order in $\epsilon$, it reads $L_{J,XPM}(\epsilon)=L_J(0)(1+4\sin^2(\frac{ka}{2})\epsilon^2)$. \\

Exchanging the role of 1 and 2, we find that this formula holds when the strong amplitude wave propagates on the $\Sigma$ mode and the weak amplitude wave on the $\Delta$ mode. When both waves propagate on the same mode, we can expand the sine functions (whose arguments are now the sum of the phases of both waves), and again find the same expression for the cross phase-modulation. We also note that to leading order, cross phase-modulation is twice stronger than self phase-modulation, as is well-known in single-mode TWPAs~\cite{OBrien2014}. \\

 In the above reasoning, when equating terms with phases rotating as $\theta_{1,n}$ from both side of the equation of motion, we  kept  resonant terms only---in the spirit of the rotating wave approximation---assuming that $\omega_1 \neq \omega_2$. A related situation arises in our experiment when $\omega_1=\omega_2$ and $k_1=-k_2$. One may then use a similar approximation assuming that terms whose phases rotate \textit{spatially} with opposed phases will not lead to significant interactions. In practice, the velocity of the modulated wave oscillates spatially, which may marginally impact the phase-matching of target processes.

\subsection{Generalized flux across a junction}
\label{sec:phi-JJ}
We here relate the generalized flux across a junction $\Phi_{\textrm{JJ}}$, used to reference the critical pump amplitude in Fig.~\ref{fig3}d, to the reduced pump amplitude $\epsilon_P$. As in Eq.~\eqref{eq:phasedrop0}, we can re-write
\begin{equation}
\varphi_{\Delta,n+1}-\varphi_{\Delta,n}=4\epsilon_P \sin(\frac{k_P a}{2})\sin(\theta_{P,n}-\frac{k_Pa}{2})
\label{eq:phasedrop1}
\end{equation}
Assuming that the pump wave propagates on the $\Delta$ mode only, we then get the peak amplitude of the flux across any single junction
\begin{equation}
\Phi_{\textrm{JJ}}=4\varphi_0\epsilon_P \sin(\frac{k_P a}{2})
\label{eq:phasedrop2}
\end{equation}
We note that in the calibration   used in Fig.~\ref{fig3}d, we account for pump reflection by the defective junction. $\Phi_{\textrm{JJ}}$ is then the peak amplitude for the generalized flux across the junctions, resulting from the superposition of a forward and a backward propagating pump wave.

\subsection{Corrections to the phase-matching relations}

\label{sec:gap-shift}

Using the  dispersion relations~\eqref{eq:correc-lumped} and including the wave velocity renormalization due to self and cross phase-modulation, we no longer obtain closed form expressions for the frequencies of the gaps associated to the  processes investigated in this work. However, we can still express the phase-matching conditions as transcendental equations. As an example, we establish the equation for the signal to idler conversion of the isolation process.\\

Letting $\epsilon_P$ the pump amplitude (traveling on the $\Delta$ mode) and $\epsilon_{S,I}\ll \epsilon_P$ the amplitude of the signal and idler waves (traveling on the $\Sigma$ mode), we first neglect phase modulation of all waves by the signal and idler, and define $L_{J,\Sigma}=L_{J,XPM}(\epsilon_P)$ and $L_{J,\Delta}=L_{J,SPM}(\epsilon_P)$. Then,  from the energy~\eqref{eq:energy} and impulsion~\eqref{eq:momentum} conservation, we get the following transcendental equation:
\begin{multline}
    -\arccos\left( 1+\frac{C_g L_{J,\Sigma}\,(\omega_S+2\omega_P)^2}{-2+2C_J L_{J,\Sigma}\, (\omega_S+2\omega_P)^2} \right)\\=
    \arccos\left( 1+\frac{C_g L_{J,\Sigma}\,\omega_S^2}{-2+2C_J L_{J,\Sigma}\, \omega_S^2} \right)
    -2\arccos\left( 1+\frac{(C_g+2C_i) L_{J,\Delta}\omega_P^2}{-2+2C_J L_{J,\Delta}\, \omega_P^2} \right)
\label{eq:correction-all}
\end{multline}

We then numerically plot the solutions to this equations, as well as the corresponding ones for the tunable coupling and aliased circulation processes, and manually fit the gap lines on the colormaps of Fig~\ref{fig2}. The fit parameters are reported in Table~\ref{table:parameters}. These are the products $L_J C_g$, $L_J (C_g+2C_i)$, the junction plasma frequency $f_{\textrm{plasma}}=\omega_J/(2\pi)$, and the  pump amplitude referenced in terms of phase drop across a junction $\Phi_{JJ}$.

\begin{table}[h]
\centering
\begin{tabular}{|c | c | c|} 
 \hline
 Parameters & Fit & Design \\ [0.5ex] 
 \hline\hline
  $v_{\Sigma,0}=1/\sqrt{L_J C_g}$ (cell/ns) & 93.6 & 90.4 \\ 
 \hline
 $v_{\Delta,0}=1/\sqrt{L_J\left( C_g + 2C_i \right)}$ (cell/ns) & 30.15 & 28.9 \\
 \hline
 $f_{\textrm{plasma}}$ (GHz) & 32.9 & /\\
 \hline
  $Z_{\Sigma}=\sqrt{L_J/C_g}$ & / & $89~\Omega$\\
 \hline
 $\Phi_{JJ}~(\Phi_0)$ & 0.12 & /\\
 \hline
 \end{tabular}
 \caption{Fit and design parameters of the device.}
 \label{table:parameters}
\end{table}

Note that these parameters are linked to the speed of the waves, while the parameters $L_J$, $C_g$ and $C_i$ do not appear isolated in the transcendental equations. To extract the individual parameters, an additional measurement is required. Impedance measurement would be ideal, as $Z_{\Sigma}=\sqrt{\frac{L_J}{C_g}}$. It is accessible through time domain reflectometry. However, in our case this proves to be too unprecise for quantitative estimation, as developed in supplemental section~\ref{sec:tdr}.

\section{Feedlines calibration and scattering by the defective junction }
\label{sec:amplitudecalibration}

Our experimental setup presented Sec.~\ref{sec:setup} features a switch and microwave standards. This allows  to measure our system scattering parameters referenced at the input port of the hybrid couplers addressing the two propagation modes. This calibration does not allow us to distinguish  losses in the microwave components and cables between the switch and the device from losses within the device itself. From an operational perspective, it still gives an accurate description of the device performances if one were to include it in a circuitQED experiment. However, this calibration does not allow us to relate the amplitude of microwaves traveling within the device to their voltage at room temperature as this requires to calibrate separately the attenuation of input and output feedlines. In order to do so, we need an \textit{in situ} calibration of the waves amplitude.\\

In the previous section, we showed how a wave of large  amplitude applied on either the $\Sigma$ or $\Delta$ mode leads, through phase modulation, to the lowering of its own propagation velocity and that of other waves with weaker amplitude. A reduced phase velocity can easily be detected by recording the phase of transmitted waves. We use such measurements to estimate the amplitude of waves within our device. \\

However, when doing so and assuming that a microwave applied from one port propagates in a well-defined direction along a single mode of the device leads to a discrepancy. Indeed, after calibrating the wave amplitude through phase-modulation, we can infer the attenuation of feedlines (up to fictitious ports directly addressing the line modes) and estimate the device intrinsic scattering parameters. We then observe significant wave reflection and inter-mode leakage, at odds with our working hypothesis.\\

These syndromes indicate that a defect within our device leads to significant wave scattering. We locate this defect close to the middle of the device through time domain reflectometry, as detailed in Sec.~\ref{sec:tdr}. We then consider a simple model in which one Josephson junction is defective and replaced by an open (Sec.~\ref{sec:model-open}). From there, we estimate the amplitude of waves propagating in the forward and backward direction on both modes when driving the device from a single-port. This allows us to reproduce the measured pattern of self and cross phase-modulation (Sec.~\ref{sec:SPM-XPM}). Finally, we use our model to calibrate the attenuation of the system feedlines, allowing us to extract the intrinsic scattering parameters of the device (Sec.~\ref{sec:intrinsic-scat}).

\subsection{Time domain measurement}
\label{sec:tdr}

\begin{figure}[h]
    \centering
    \includegraphics{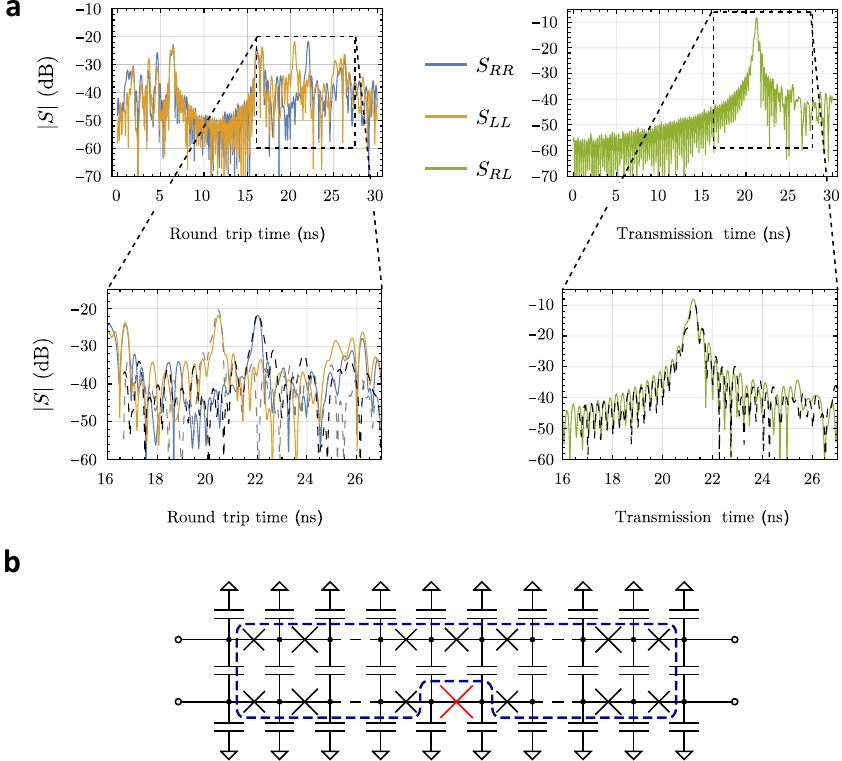}
    \caption{{\bf Time domain measurements. a:} full range of time domain impulse response measurements in reflection and transmission referenced at the $\Sigma$ ports of the hybrid couplers in Fig.~\ref{fig:setup}. We identify the portion of the time domain that corresponds to the lumped transmission line. The dip in reflection corresponding to the transmission through the 1~m cable connecting the sample is a good indicator of the relevant portion. {\bf Dashed square boxes:} magnified time domain measurements within the Josephson transmission line. The dashed lines are the time domain impulse response of a model featuring a single open junction on cell no~165 and residual disorder in other cells (see (b)). There is qualitative agreement with the data, in particular reproducing the two reflection peaks that are at a different position depending on the input direction. Exact agreement is not possible as the model is randomized, and sensitive to the specific realization of disorder. In the right panel picturing the time response in transmission, we notice that the main peak is wider than the resolution of 300~ps. This is consistent with the dispersion caused by the plasma frequency of the junctions, which is included in our model. {\bf b:}  Single open junction model used to reproduce the time domain impulse response of the sample. The red  cross represents an open junction. The blue dashed rectangles indicate that disorder is applied to the other junction, modeling the fabrication parameters spread. To model disorder, we sample the Josephson junction inductances uniformly in an interval [0.95 $L_J$, 1.05 $L_J$] centered at the nominal design value.}
    \label{fig:tdr}
\end{figure}

In order to estimate the level of reflection within the sample and  confirm the hypothesis that reflection primarily occurs within the active Josephson line itself (and not to the connecting circuitry), we perform time domain reflectometry measurements. These measurements also allow us to pinpoint the position of the reflective defect.\\

Traditionally, time domain measurement is performed by launching a short pulse or a stepwise excitation pulse towards the device under test, and monitoring as a function of time the reflected and transmitted signals. However, this is impractical and noise-prone in complex radio-frequency systems~\cite{KeysightTDR} such as our device in the cryostat. Fortunately, the time domain measurement is virtually equivalent to a frequency domain one, up to a Fourier transform~\cite{KeysightTDR}. In addition, frequency domain measurement benefits of the increased precision from the calibration. The analyzed portion of the system, and consequently the complexity of interpretation of the reflection pattern, is greatly reduced thanks to the shift of the reflection planes closer to the sample obtained using calibration standards (see section~\ref{sec:setup}).\\

Despite all these benefits, there  are still limitations. Because of the circulators, the device can only be measured in a limited band, excluding DC and low-frequencies. This has the double inconvenient of reduced resolution (as low-pass data including DC could have been mirrored to negative frequencies, doubling resolution) and  inability to reconstruct the response of a system to a step input (which necessitates DC response). Only an impulse response can be computed, for which the magnitude of the reflected and transmitted peaks is less reliable than the magnitude of the step response. Each successive reflection before the impulse reaches the two-mode Josephson line in the calibrated data adds uncertainty. Indeed the power reaching a given impedance defect, and its associated reflection magnitude that we detect, is affected by the previous reflections. Additionally, the complex phase of the impulse response is hardly usable, whereas the sign of the real step response would have been easily interpreted to distinguish types of impedance mismatches.\\

We perform the inverse Fourier transform on frequency domain S-parameters of the $\Sigma$ mode, while no pump is being applied to the $\Delta$ mode, to obtain the time domain S-parameters in Fig~\ref{fig:tdr}. The original frequency data is sampled from 4 GHz to 8 GHz, that is a 4 GHz bandwidth. With minimal windowing applied, this results in a resolution of about 300 ps (1.2/BW, \cite{KeysightTDR}). In a typical coaxial cable in which waves propagate at $0.7c$, the resolution is about $6\cdot10^{-2}$~m. However, in the non-linear device, speed~\eqref{eq:speeds} is 93.6~cell/ns, yielding a resolution of 28 cells, enough to approximately locate a defect within the line. The results of this effective time domain measurement is presented in Fig.~\ref{fig:tdr}, and exhibits multiple reflection peaks separated by periods where the reflected signal is negligible. We interpret the long dip in reflection as the 1~m cable connected to the sample (expected 9.5~ns for the round-way trip), and identify the following portion as the device itself (only the lumped element line part is resolvable given the greatly reduced speed inside it). In dashed square boxes in Fig~\ref{fig:tdr}, the relevant portion is magnified. Two reflection peaks are noticeable close to the middle of the transmission line. This corresponds to internal  reflections by a scattering defect. The reflection peaks for impulses applied from  each input port do not overlap as the defect is not perfectly centered in the line. We reproduce qualitatively the reflection response using a two-mode model (dashed lines in Fig~\ref{fig:tdr}b) where a single open junction is responsible for most of the reflection (see Sec.~\ref{sec:model-open} for details). To model disorder, we  sample the inductances of other Josephson junctions uniformly in an interval [0.95 $L_J$, 1.05 $L_J$] centered at the nominal design value. This allows us to reproduce the reflection floor. \\

With this model and analyzing reflection peaks when the nonlinear two-mode line connects to the tapered feedline (designed such that negligible reflections should occur if the nonlinear portion is perfectly on specification), we can estimate the impedance of the $\Sigma$ propagating mode on the order of $100~\Omega$, in qualitative agreement with the design value. However, the quantitative results are not reliable because of the previously explained time-domain technique limitations.

\subsection{Single open junction model}
\label{sec:model-open}

After considering several types of defects (open junction, shorted junction, shorted capacitor), we opt for a single open junction model as pictured in Fig.\ref{fig:tdr}b. This model allows us to qualitatively capture all signatures of wave scattering pictured in Fig.~\ref{fig:scattering}.\\

\begin{figure}[h]
    \centering
    \includegraphics[]{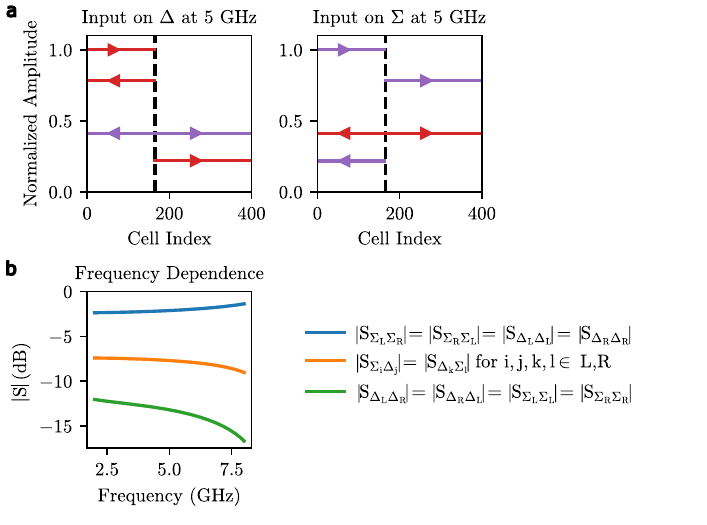}
    \caption{ {\bf Wave scattering by an open junction. a:} Simulated amplitude profile of forward and backward propagating waves, along both modes of the device, in presence of a defect (open junction) located on cell 165. The amplitudes are referenced by that of the applied wave, which is small enough that non-linear effects can be neglected. The line parameters are those reported in Tab.~\ref{table:parameters}, and the wave frequency is 5 GHz.  Left panel: wave applied from the left $\Sigma$ port. Right panel: wave applied from the left $\Delta$ port. {\bf b:} Fraction of the  wave power transmitted, reflected and leaked to the other mode when scattered by the defect, referenced to the input wave amplitude,  as a function of the input wave frequency. Our model is only applicable below the $\Delta$ mode cutoff frequency at $9.44$~GHz.
    }
    \label{fig:waveprofiles}
\end{figure}

In Fig.~\ref{fig:waveprofiles} we represent the amplitude profiles of waves scattered by the open junction in both directions and on both modes of the device. Noticeably, the defect does not affect waves applied from the $\Sigma$ port or $\Delta$ port in the same manner. For our line parameters, about 60~\% of the power of a wave traveling on the $\Sigma$ mode is transmitted through the defect, whereas the same fraction is reflected for a wave traveling on the $\Delta$ mode. In both cases, about 20~\% of the wave power leaks in each direction on the other mode. Only 5~\% of the wave power is transmitted on the $\Delta$ mode and the same fraction is reflected on the $\Sigma$ mode.\\

This has important consequences for our device performance. In particular, in the circulation process, the strong reflection of pump waves (on the $\Delta$ mode) causes reverse isolation that increases with pump amplitude, limiting the device maximum isolation (see Fig.~\ref{fig3}). Moreover, due to the small fraction of transmitted pump power, the section of the Josephson line downstream of the defect remains inactive. This effectively halves the device's electrical length and significantly limits its performance. Finally, strong inter-mode leakage is expected, which would negate any benefit gained from spatially separating the pump and signal waves if our device were to be interfaced with sensitive quantum systems.

\subsection{Self and cross phase-modulation measurements}
\label{sec:SPM-XPM}

\begin{figure}[h]
    \centering
    \includegraphics[]{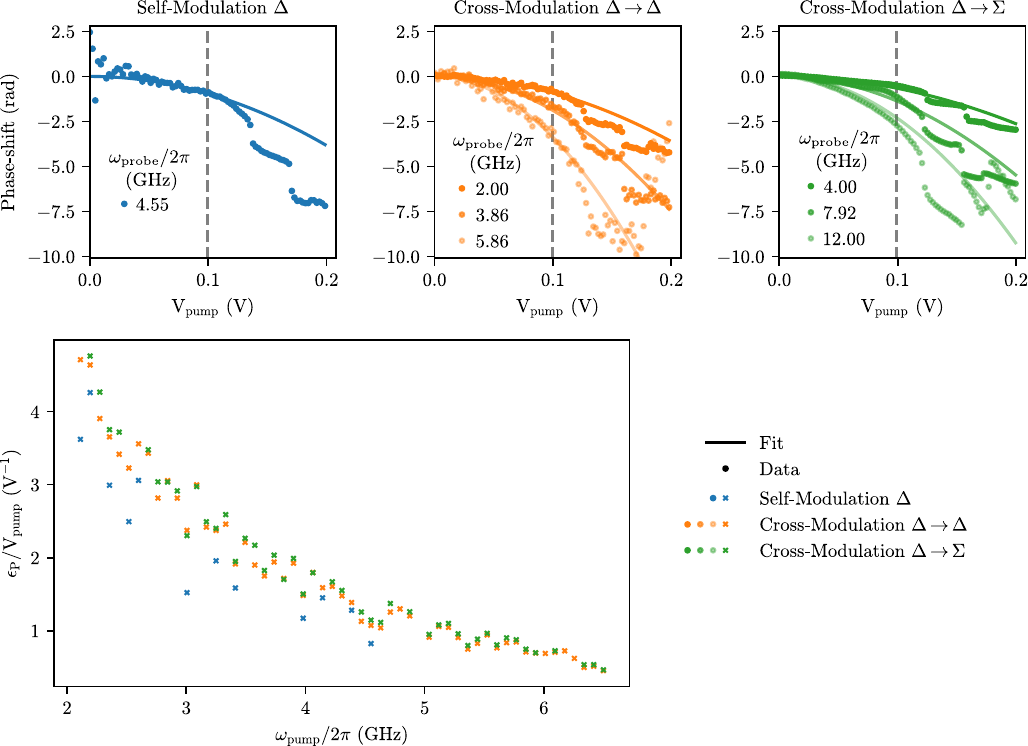}
    \caption{ {\bf Input lines calibration. Top panels:} For  a pump at 4.55~GHz applied on a $\Delta$ port of the device, we record the phase-shift of the transmitted wave induced by self phase-modulation (left panel),  as a function of the pump voltage at room temperature. We subsequently record the phase-shift induced by cross phase-modulation on an additional probe wave transmitted on the $\Delta$ mode (middle panel) or on the $\Sigma$ mode (right panel), for a varying probe frequency (encoded in the shade of marker color).  Our model (see text for details) is fitted to each set of curve independently, yielding
    three independent estimates for the ratio between the pump voltage at room temperature and its amplitude within the line. We only consider phase-shifts induced by low-amplitude pumps (below the amplitude marked by dashed vertical lines) in our analysis. Parabolic lines are results from the fits. We attribute discontinuities in the measured phase-shifts at larger pump amplitude to the interplay between the line non-linearity and internal pump reflections. {\bf Bottom panel:} For each pump frequency considered in this work, we report the three independent estimates between the pump voltage at room temperature and its amplitude within the line (referenced in terms of generalized flux across junctions located right of the defective junction). The two estimates from cross-modulation are in quantitative agreement and differ only slightly from estimates based on self-modulation. The latter are less reliable as they are based on phase-shifts detected at a single pump frequency only, and cannot always be extracted (missing blue points).
    }
    \label{fig:SPM-XPM}
\end{figure}

With the model described in the previous section, we can predict, for an input wave applied with a given amplitude at a given port, the amplitude of waves traveling in both directions and on both modes within the device. Using Eqs.~\eqref{eq:SPM}, \eqref{eq:XPM}, we  then estimate the  phase-shift of the transmitted wave due to self phase-modulation, along with the phase-shift induced on a second, low-amplitude wave due to cross phase-modulation. Comparing the predicted phase-shifts with the experimentally measured ones, we obtain the ratio of the strong  wave amplitude within the device to its voltage at room temperature, corresponding to the input lines attenuation.    Fig.~\ref{fig:SPM-XPM} reports the results of this calibration for the input line connecting to one of the $\Delta$ ports of the device (in this context, $\Sigma$ and $\Delta$ "ports" should be understood as referring to fictitious ports addressing either mode of the device from either end of the Josephson line). We note that  our model consistently captures the phase-shift pattern induced by self-phase modulation of a strong wave and by cross phase-modulation on a second, low-amplitude probe over a wide range of probe frequencies. We perform similar measurements to calibrate the attenuation of the three other input lines (not shown).

\subsection{Intrinsic scattering parameters}
\label{sec:intrinsic-scat}

\begin{figure}[h]
    \centering
    \includegraphics[width=1\columnwidth]{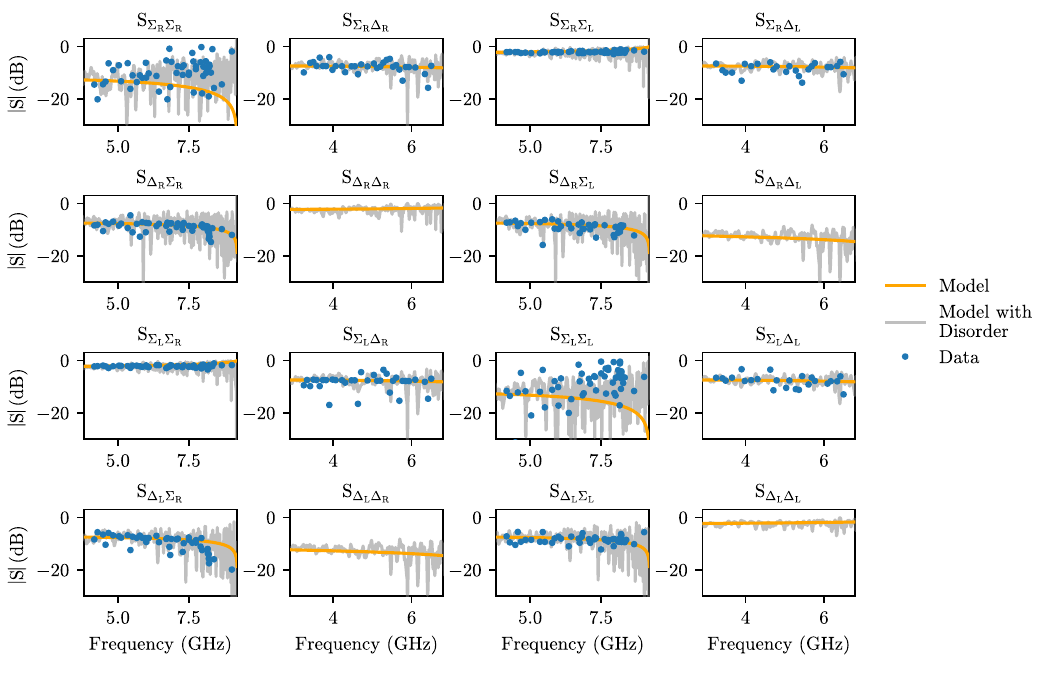}
    \caption{{\bf Calibrated scattering parameters of the device. } After calibrating the attenuation of the system input lines (see \ref{sec:SPM-XPM}), we fit our model to the device scattering parameters (measured at room temperature) to extract the attenuation of output lines (4 free fit parameters). We report here the scattering parameters after correcting for both the input and output lines attenuation. Blue dots are the measured scattering coefficients. Reflection coefficients on $\Delta$ ports and transmission coefficients between one $\Delta$ port and the other could not be reliably extracted (see text). Orange lines represent the predictions from our model featuring a single open junction, and in which the device is probed from impedance-matched $\Sigma$ and $\Delta$ ports. Gray lines represent the predictions from a refined model in which the inductances of all other junctions  are randomly sampled in an  interval [0.95 $L_J$, 1.05 $L_J$] around their average value, and the device is probed via  ports  with  impedances $\tilde{Z}_{\Sigma}=89~\Omega$ and $\tilde{Z}_{\Delta}=28~\Omega$ corresponding to the matching impedances of the tapered adapter (see Sec.~\ref{sec:layout}). 
    }
    \label{fig:scattering}
\end{figure}

Having calibrated the attenuation of the system input lines, we self-consistently validate our model featuring a single open-junction by reconstructing the system scattering coefficients referenced at  fictitious ports addressing the two modes of the line (correcting for the attenuation of input and output lines).\\

To do so, we first measure the system scattering coefficients referenced at room temperature. We are able to reliably measure 12 such coefficients. The four missing coefficients  are the reflection coefficients  on both $\Delta$ ports of the system and the transmission coefficients from one $\Delta$ port to the other. Indeed, in these configurations, probe waves are heavily attenuated passing twice through a -30~dB attenuator (see Fig.~\ref{fig:setup}) such that waves leaking through the directional couplers preceding the attenuator (in the former case) or through our room temperature setup (in the latter case) dominate over those following the probed path.\\

Second, at each probe frequency, we fit the magnitude of the 12 coefficients with our model, leaving as free fit parameters the  attenuation of each of the 4 output lines of the system~\footnote{At a few probe frequencies, the transmission of the input lines could not be extracted due to strong ripple in the phase-modulation measurements. In these cases, the input line attenuation is left as a free fit parameter, resulting in a fit that is not over-constrained for some panels (blue dots lying exactly on the orange line.}. We repeat this procedure over the frequency range in which our model is valid (limited by the  cut-off frequency of the $\Delta$ mode at 9.44~GHz) and by the working bandwidth of the circulators placed on $\Sigma$ output lines. We then compare the scattering coefficients referenced at the device fictitious ports (blue dots in Fig.~\ref{fig:scattering}) to the prediction from our model (orange line). Predictions from our model qualitatively matches the recorded pattern, but does not account for the dispersion of the measured coefficients around their average value.\\

 We attribute this dispersion to  disorder in the properties of the line unit cells and impedance mismatch at the ports.  Enriching our model by sampling the value of the inductances of Josephson junction inductances  in an interval [0.95 $L_J$, 1.05 $L_J$] centered around their average value $L_J$ (same model as used in Fig.~\ref{fig:tdr}) and setting the ports impedance to the values of the tapered adapter (see Sec.~\ref{sec:layout}), we predict a dispersion of the scattering coefficients on the same scale as observed  experimentally. \\

Several  remarks are in order concerning the calibration presented in this section.
\begin{itemize}
\item Our model does not include dielectric loss in the line nor any other source of dissipation. In Sec.~ \ref{sec:setup}, we detailed how we rigorously calibrated the scattering parameters of the device referenced at the ports of the hybrid couplers. In a separate cooldown, we measure the transmission through the hybrid couplers and in-line cables and filters after replacing the device with two short coaxial cables. We thus estimate these components to be responsible for 2.5 to 6.5 dB of loss. Subtracting this loss as well as apparent loss induced by scattering on the defective junction from the attenuation measured between the two $\Sigma$ ports of the hybrid couplers, we estimate dissipative loss within the Josephson line to range from 1 to 2.5~dB (consistent with tangent losses $\tan\delta\approx2\cdot10^{-3}$ in the capacitors's alumina). Missing to include this loss in our model sets a bound on the precision of our calibration.
\item The device intrinsic scattering matrix estimated in this section should be reciprocal (in absence of a pump). In Fig.~\ref{fig:scattering}, forward and backward transmission coefficients differ by up to  3~dB, confirming the level of accuracy of our calibration.
\item In Fig.~\ref{fig:scattering}, our model fails to capture an increase of the magnitude of the reflection coefficients on the $\Sigma$ input ports when approaching the $\Delta$ mode cutoff frequency at 9.44~GHz. In any case, our model is only applicable below the cutoff frequency, limiting the frequency range of our calibration.
\end{itemize}
As a conclusion, the methods presented in this section  allow us to get a coarse absolute calibration of wave power within the device, which cannot be inferred with the switch-based method presented in Sec.~ \ref{sec:setup}. This calibration allows us  to validate our model featuring a single open junction within the line, to estimate the device dynamic range in Sec.~\ref{sec:dyn-range} and to coarsely verify photon number conservation in the conversion process in Sec.~\ref{sec:conversion}. However, this calibration is too imprecise to estimate the circulation efficiency of our device.

\section{Detailed performance in circulation}

In this section, we present supplemental data pertaining to the circulation process and compare experimental data with the predictions of a simple analytical model. We further compare these predictions with those from a more refined model accounting for the discrete structure  of the line and  a larger number of active processes. We solve the latter model numerically following the method introduced in Ref.~\cite{peng2022x}.\\

\subsection{Accuracy of the analytical model}
\label{sec:continuous-vs-full}
The assumptions and calibrations  underlying the analytical model are the following.
\begin{enumerate}
\item The line is treated as a continuous medium and we assume the  amplitude of waves to vary slowly compared to their wavelengths. We only account for the discrete structure of the line by adjusting the line dispersion relation (through Eq.~\eqref{eq:correc-lumped}).
\item The pump wave (frequency $\omega_P$) is stiff and a single conversion process between the signal at $\omega_S$ and the idler at $\omega_S+2\omega_P$ is active. We thus neglect pump harmonic generation (see \ref{sec:pump-harmonics}) and conversion of the signal to other sidebands at $\omega_S+2n\omega_P$ with $n\neq 1$. Beside this conversion process, the only consequence of the  line   non-linearity which we account for is a renormalization of the dispersion relation set by the pump amplitude (through Eq.~\eqref{eq:correction-all}).
\item An open junction near the middle of the line leads to strong scattering of waves (scattering due to disorder within the line and at the ports are neglected). The resulting amplitude  profile for the pump wave is depicted in Fig.~\ref{fig:waveprofiles} (left panel).  We account for the appearance of a reflected  pump and for the depletion of the transmitted pump  on the $\Delta$ mode. After estimating the pump amplitude in each direction and on each side of the defect, we solve the signal and idler wave dynamics on the $\Sigma$ mode, before and after the defect, with joint boundary conditions set by the defect.  The pump wave leaked to the $\Sigma$ mode is neglected. Scattering of signal and idler waves by the defect contribute to the recorded device insertion loss, corresponding to the attenuation at zero pump amplitude in Fig.~\ref{fig3}b and Fig.~\ref{fig:extended-data}. 
\item The pump amplitude referenced at the device input ports is calibrated following the method detailed in Sec.~\ref{sec:amplitudecalibration}, with a $\sim 3$~dB precision.
\end{enumerate}
Note that (1) and (2) are closely related to the assumptions made in Sec.~\ref{sec:bands}, provided that the pump and signal frequencies are adjusted to phase-match the circulation process. The only difference is that we account for corrections to the line dispersion relation stemming from the discrete structure of the line and from phase modulation. (3) and (4) arise from the model and calibration described in Sec.~\ref{sec:amplitudecalibration}.\\

\begin{figure}[h]
    \centering
    \includegraphics[width=0.7\columnwidth]{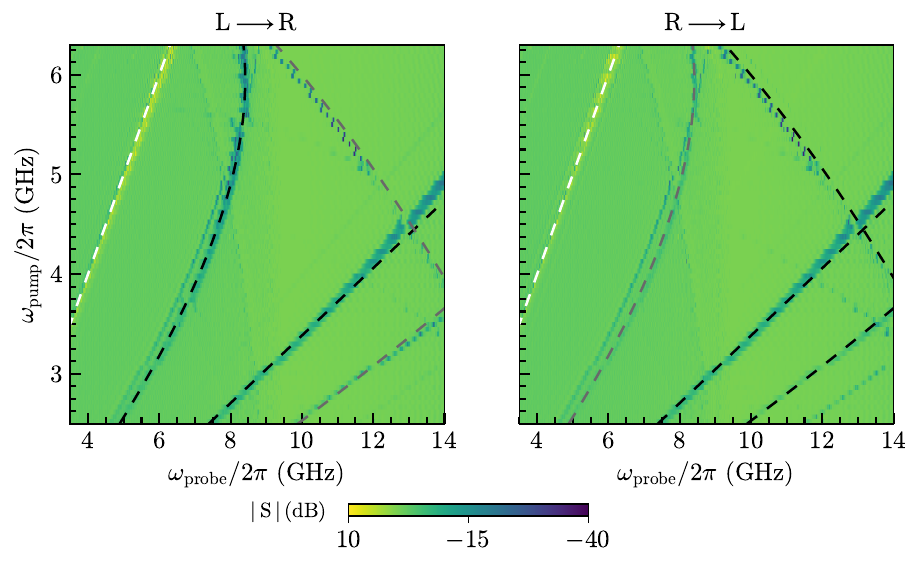}
    \caption{{\bf Forward and backward transmission in the non-linear discrete (NLD) model}  We solve numerically~\cite{peng2022x} the dynamics encoded by the Lagrangian~\eqref{eq:Lagragian-sym}, with parameters fitted to the analytical model and including an open junction (see text for details). This allows us to predict the $\Sigma$ mode transmission for varying pump and probe frequencies and a set pump amplitude as in Fig.~\ref{fig2}. Simulations reproduce experimental data up to an offset in background transmission (insertion loss not included in the model, colorscale offset by 10~dB compared to the one used in Fig.~\ref{fig2}), confirming the accuracy of the estimated system parameters. Overlaid black and gray dashed lines represent the modified phase-matching relations for the circulation process (Eq.~\eqref{eq:correction-all}), the alias process and the coupling process. The white dashed line marks amplification at $\omega_{\mathrm{probe}}=\omega_{\mathrm{pump}}$
    }
    \label{fig:sim-colormap}
\end{figure}

In order to test the accuracy of the analytical model, we compare its predictions with those from a non-linear, discrete model (referred to as the \textit{NLD model}, corresponding to the Lagrangian~\eqref{eq:Lagragian-sym}), solved by adapting the methods introduced in Ref.~\cite{peng2022x}. The numerical simulations are based on the open-source library JosephsonCircuits.jl written by the authors of this reference and use a two-step harmonic balance to (1) compute the generalized flux across each junction induced by the pump and (2) compute the signal wave X-parameters after renormalizing the inductance of each junction owing to the pump wave. \\

We first test the accuracy of the modified phase-matching relation~\eqref{eq:correction-all} used in the analytical model with NLD simulations of the $\Sigma$ mode transmission against pump and probe frequency. These simulations are made in the  conditions in which the data presented in Fig.~\ref{fig2} are acquired, and with system parameters fitted  to the analytical model (see Table.~\ref{table:parameters}). For the missing parameter---setting the impedance of the device modes---we use the design value for lack of a precise calibration method. Moreover, we include in simulations an open junction at position 165 as found in Sec.~\ref{sec:amplitudecalibration}. In Fig.~\ref{fig:sim-colormap}, we report the simulated  transmission. Overlaid black and gray dashed lines denote the predicted pairs of frequencies $(\omega_{\mathrm{probe}},\omega_{\mathrm{pump}})$ at which the analytical model predicts a phase-matched process. Simulation results are in good agreement with the recorded data presented in Fig.~\ref{fig2}, up to an offset in background transmission as several sources of insertion loss present in our experiment  are not featured in the simulations (loss in hybrid couplers and cables, dielectric loss). This confirms that the values of parameters estimated from the analytical model, as well as the modified phase-matching relations,  are accurate. \\

The NLD simulations also reproduce the peak at $\omega_{\mathrm{probe}}=\omega_{\mathrm{pump}}$, revealing amplification. This feature could  have also been captured by the analytical model, but was not included in our analysis so far. Indeed, the corresponding process destroys two counter-propagating photons at $\omega_{\mathrm{pump}}$ on the $\Delta$ mode and creates two counter-propagating photons at $\omega_{\mathrm{probe}}=\omega_{\mathrm{pump}}$ on the $\Sigma$ mode, preserving energy and momentum (equations \eqref{eq:energy} and \eqref{eq:momentum}). Note that since the signal photon and the corresponding idler photon propagate in opposite directions, we expect amplification to be phase-preserving, even though both are at the same frequency. Moreover, the reversed direction of propagation for the idler is expected to yield a different dynamics compared to regular amplification in TWPAs~\cite{OBrien2014}, which will be investigated in a future work. \\

The NLD simulations also feature a narrow  gap at a slightly lower frequency than the circulation gap  (visible left of the gray dashed line on the right panel). We are so far unable to identify the corresponding mechanism, which seems to involve waves leaking from one mode to the other since the gap disappears when performing similar simulations without a defective junction (even applying a pump from both $\Delta$ ports).\\ 

\begin{figure}[h]
    \centering
    \includegraphics[width=0.8\columnwidth]{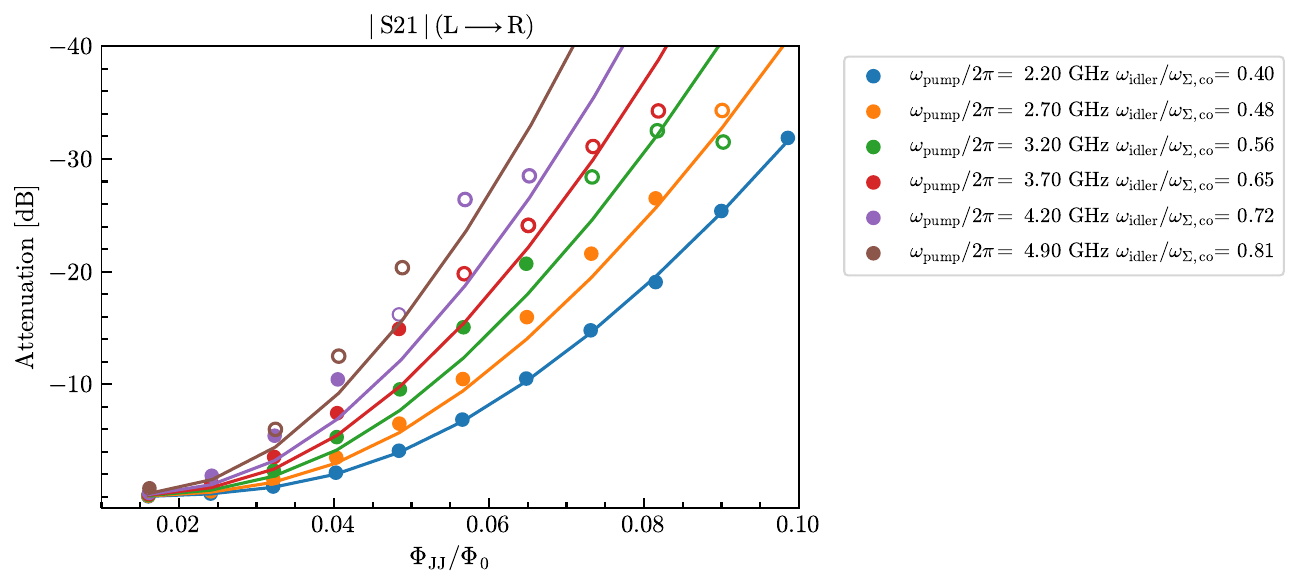}
    \caption{{\bf Validity of the continuous model for a uniform line. } We compare the isolation predicted by the analytical model (lines) with that predicted by the NLD model for various pump amplitudes (horizontal axis, referenced in terms of generalized flux across a junction) and pump frequencies (encoded in color). We \textit{do not} include a defective junction in either model. The probe frequency is always adjusted at the center of the circulation gap. Predictions from the NLD model deviate from those of the analytical model, which is based on a low-frequency approximation, as the idler frequency approaches the $\Sigma$ mode cutoff frequency (see Eq.~\eqref{eq:signal-freq}). For the highest pump frequencies and amplitudes,  the circulation gap predicted by NLD simulations features a flat bottom and/or strong irregularities, which we indicate by empty circle markers (not shown). In these cases, in-band attenuation is estimated by smoothing the frequency response. Missing markers indicate that NLD simulations failed altogether (missed harmonic balance or chaotic frequency response).
    }
    \label{fig:continuous-discrete}
\end{figure}

Next, we  test the domain of validity of the hypotheses (1) and (2) (continuous model and single active process, implying in particular that the pump is stiff). In Fig.~\ref{fig:continuous-discrete},  we compare the  isolation predicted by the analytical model to that  predicted by NLD simulations, for various pump frequencies and amplitudes, the probe frequency being adjusted such that the circulation process is phase-matched. Here, for simplicity, we do not include a defective junction in either model (all cells are identical). \\

At low pump frequency, predictions from the two model match quantitatively. As the pump frequency increases and the corresponding idler frequency (set by Eq.~\eqref{eq:signal-freq}) approaches the $\Sigma$ mode cutoff frequency, predictions from both models start deviating from one another. This is expected given that the analytical model is based on a low-frequency assumption, allowing modeling of the line as a continuous medium. Nevertheless, predictions from the NLD model do not qualitatively deviate  from those of the analytical model in the low pump amplitude regime. In particular, for the pump frequency and amplitudes considered in Fig.~\ref{fig:continuous-discrete}, the analytical model under-predicts the attenuation by up to 8~dB, but this imprecision is not significant compared to that entailed by the finite accuracy with which we calibrate the pump amplitude and to the unpredictability entailed by the defective junction (see Fig.~\ref{fig:extended-data}). At higher pump amplitudes, the NLD simulations fail, either due to missed harmonic balance or chaotic frequency response (not shown).\\

As a concluding remark, we note that in the case that we consider here of a uniform line with no defect,  the pump amplitude $\Phi_{JJ}$ corresponds to a forward propagating pump only, whereas in presence of a defect, it corresponds to the peak amplitude of the sum of a forward and a backward propagating pump. Moreover, in presence of a defect in the middle of the line, only the line section prior to the defect is effectively active (see Sec.~\ref{sec:amplitudecalibration}). These key differences explain the larger attenuations predicted in the present case compared to those measured experimentally.

\subsection{Isolation and insertion loss}
\label{sec:extended-atten}
\begin{figure}[h]
    \centering
    \includegraphics[]{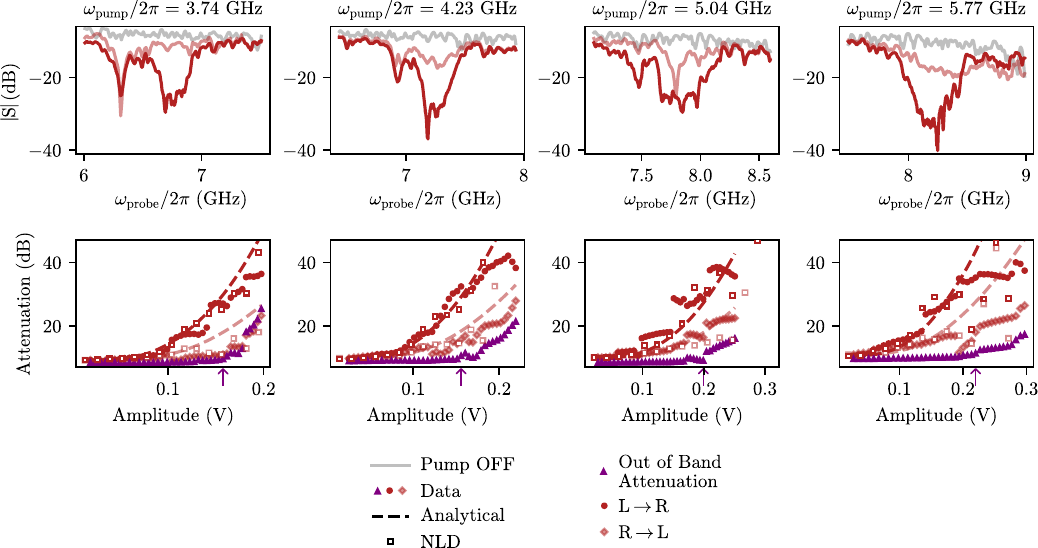}
    \caption{{\bf Circulator performance in isolation}. We present supplemental data for the circulation process analog to those displayed in Fig.~\ref{fig3}b. {\bf Bottom panels:} at four selected pump frequencies, we report the $\Sigma$ mode transmission in the forward (dark red dots) and backward (light red diamonds) direction, at the center of the circulation gap, against pump amplitude. We further show the forward attenuation out of the circulation band defined as the average attenuation over two 1~GHz-wide frequency bands located on either side of the gap (purple triangles). Beyond critical pump amplitude (vertical arrows), the out-of-band attenuation increases sharply, indicating that spurious non phase-matched processes become active. Dashed lines are predictions from the analytical model. Empty square markers  are predictions from the NLD model (offset to match the measured attenuation at zero pump amplitude). Both models include a defective junction. The latter are irregular with fluctuations on the same magnitude as the experimental data. They however fail to quantitatively reproduce experimental results due to high sensitivity to simulation input parameters. {\bf Top panels:} Transmission in both directions against probe frequency and at critical pump  amplitude. 
    }
    \label{fig:extended-data}
\end{figure}

In Fig.~\ref{fig3}b, we compare the measured isolation in the circulation band, for a pump at $\omega_P/(2\pi)=4.2~$GHz,  with the predictions from the analytical model presented in Sec.~\ref{sec:bands}. We present in Fig.~\ref{fig:extended-data} supplemental data at the same pump frequency, along with three  other pump frequencies. We further report the measured probe attenuation from the left to the right $\Sigma$ port out of the working bandwidth of the circulation process. This out-of-band attenuation is defined as the average attenuation over two 1~GHz-wide frequency bands located on either side of the gap (attenuation referenced, as for the in-band attenuation, at the ports of the hybrid couplers in Fig.~\ref{fig:setup}). The analytical model qualitatively captures the recorded attenuations up to a critical pump amplitude, at which point we observe a sharp increase of the out-of-band attenuation, which is not expected and may indicate that spurious non phase-matched processes become active. Quantitatively, we define the critical amplitude as the point at which the out-of-band attenuation has increased by 2~dB above its value when the pump is off.\\

We further compare the measured attenuations  to the predictions of the NLD model. Predictions from NLD simulations are irregular with fluctuations on the same magnitude as the experimental data but fail to quantitatively reproduce the results due to high sensitivity to simulation input parameters. NLD simulations do not predict a wideband collapse of probe transmission beyond critical amplitude, but generally tend to become unreliable at large pump amplitude (see Fig.~\ref{fig:continuous-discrete}).

\subsection{Bandwidth}
\label{sec:exp-bandwidth}

\begin{figure}
    \centering
    \includegraphics[width=1\columnwidth]{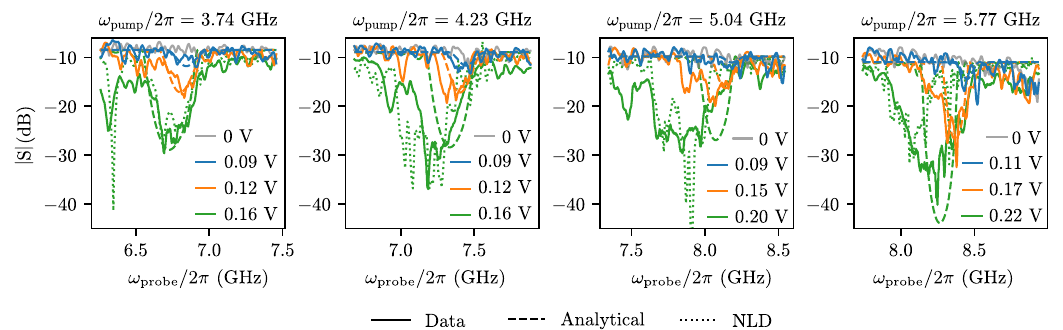}
    \caption{{\bf Frequency response of the device for the circulation process.}  We present supplemental data for the circulation process analog to those displayed in Fig.~\ref{fig3}a. Pump amplitude is encoded in color and each panel represents a different pump frequency. Predictions from the analytical model (dashed lines) are in good agreement with the data at low pump amplitude, but fail to capture the gap broadening and frequency shift observed at higher amplitude for some pump frequencies. Predictions from the NLD model (only shown as dotted lines for the highest pump amplitudes) feature similar behavior,   but fail to quantitatively reproduce experimental results due to high sensitivity to simulation input parameters. Both models include a defective junction. }
    \label{fig:detuned}
\end{figure}

We further compare the predictions of the analytical and NLD models to measured data in terms of bandwidth of the circulation process. In Fig.~\ref{fig:detuned}, we represent the  transmission, in the forward direction, of a probe in the vicinity of the circulation gap, for a few selected pump frequencies and amplitudes. Predictions from the analytical model are in good agreement with the measured data at the lowest pump amplitudes, but  fail to capture the gap broadening and frequency shift observed at higher amplitude for some pump frequencies. Predictions from NLD simulations qualitatively capture most features observed experimentally such as irregularities and unexpected broadening and shift of the gaps. However, these features occur unpredictably and are highly sensitive to simulation input parameters so that quantitative agreement is not reached. 

\begin{figure}
    \centering
    \includegraphics[width=0.5\columnwidth]{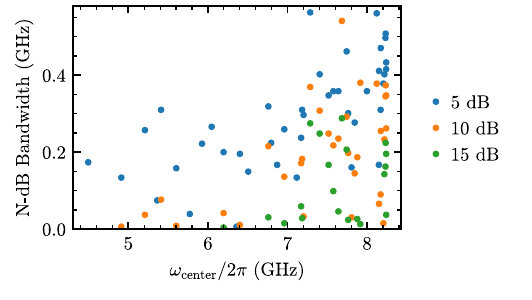}
    \caption{ Bandwidth of the circulating process as a function of center gap frequency, for a set a isolation levels. For each pump frequency, we smooth the probe transmission spectrum with a 40~MHz window and locate the two points framing the $\left\vert S_{RL}/S_{LR}\right\vert$ gap at the N-dB level (N being 5, 10 or 15). The distance between these two points is the N-dB level bandwidth.}
    \label{fig:all-bw}
\end{figure}

In Fig.~\ref{fig:all-bw}, we report the device working bandwidth in circulation mode, for a set of isolation levels and at critical pump amplitude, against the circulator gap center frequency.   For each pump frequency considered in this work, we locate the circulation gap center frequency and find the frequency range for which the non-reciprocity $-20\log\left\vert S_{RL}/S_{LR}\right\vert$ is larger than the set level. The 10-dB level is typically reached over 200~MHz for center gap frequencies above 7~GHz, and the 5-dB level over 200~MHz for most considered frequencies. \\


\subsection{Dynamic range}
\label{sec:dyn-range}
\begin{figure}[h]
    \centering
    \includegraphics[]{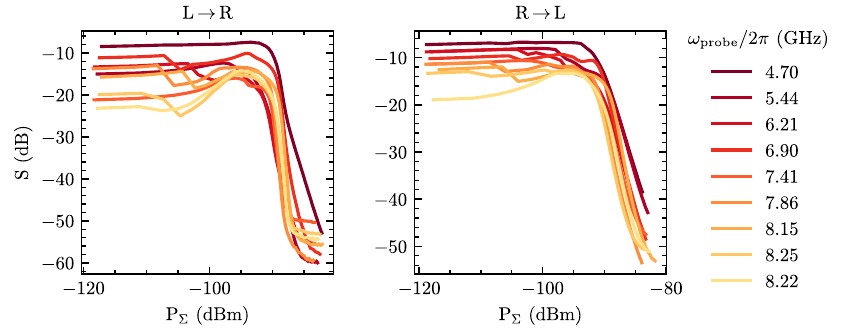}
    \caption{{\bf Dynamic range of the circulator process at various working frequencies.} The displayed frequencies correspond to linear spacing in pump frequency, the pump amplitude is adjusted at critical value and the probe power is swept (power referenced at the device input port using the calibration described in Sec.~\ref{sec:amplitudecalibration}). {\bf a,} Forward transmission. {\bf b,} Backward transmission. In both cases, transmission  is approximately constant up to a probe power of about $-110$~dBm. It can then evolve up or down depending on probe frequency, and finally collapses above $-90$~dBm.}
    \label{fig:dyn-range}
\end{figure}

In this section, we estimate the dynamic range of the device used in circulation mode, that is the maximum signal wave power it can handle before its performances degrade. For a set of pump frequencies and for pump amplitudes adjusted at critical value, we locate the center gap frequency at low probe power. We then record the forward and backward probe transmission as a function of input probe power (see Fig.~\ref{fig:dyn-range}). The isolation (defined as the ratio of backward to forward transmission) is stable up to a probe power of $-110$~dBm.  Beyond $-90$~dBm, the transmission in both directions collapses. The corresponding probe amplitude approximately matches the pump amplitude at critical value.
 \\

In a future experiment, the dynamic range could be increased by increasing the critical current of  Josephson junctions. This can either be done by adding cells to the line and reducing the values of $L_J$ and capacitances accordingly (in order to maintain a mode impedance close to 50~$\Omega$), or by keeping the total inductance per cell constant  but replacing each junction with a short chain of junctions  as done for example in~\cite{White2015}.

\newpage
\section{Power circulation, sideband generation and inter-mode leakage}

\subsection{Power circulation and inter-mode leakage}
\label{sec:conversion}

In this section, we investigate the conversion aspect of the circulation process. To proceed, we apply a pump wave with frequency $\omega_P$ and varying amplitude on the right $\Delta$ port of the device, and a low-amplitude probe wave, at frequency $\omega_{\mathrm{probe}}$ adjusted to phase-match the circulation process, on one $\Sigma$ port of the device. In practice, that means that $\omega_{\mathrm{probe}}=\omega_S$ when probing from the left port (signal band) or $\omega_{\mathrm{probe}}=\omega_I$ when probing from the right port (idler band), where $\omega_S$ and $\omega_I$ are set by the pump frequency following Eq.~\eqref{eq:signal-freq}. We then detect the power spectrum transmitted or reflected at each $\Sigma$ port. We reference input and output powers at fictitious ports directly addressing the $\Sigma$ mode of the device using the calibration methods described in Sec.~\ref{sec:amplitudecalibration}.\\

Owing to the wave mixing properties of the Josephson junction and the symmetries of the line Lagrangian \eqref{eq:Lagragian-sym}, we expect the device to generate sidebands at $\omega_{B_n}=\omega_{S} + n\omega_P$  (with $n\in 2\mathbb{Z}$) on the $\Sigma$ mode (we analyze the generation of harmonics of the pump in the next section). In particular, the reflected sideband at $\omega_I=\omega_{B_{2}}$ should be the dominant sideband when the probe is applied at $\omega_S=\omega_{B_0}$ from the left port (respectively the reflected sideband at $\omega_S=\omega_{B_0}$ when the probe is applied  at $\omega_I=\omega_{B_{2}}$ from the right port) as this conversion process is phase-matched. Nevertheless, other sidebands are expected to leak from both ports  due to spurious unmatched  processes.\\

In practice detecting a large number of sidebands is complex. Indeed, the pump frequencies are comparable to the signal frequencies, and the sidebands spread from near-DC to the cut-off of the line, over 20 GHz. We only measure sidebands that are within the isolators and amplifiers bandwidth, whose low-frequency limit is at 4~GHz. Moreover, our calibration for the system feedlines is only valid below the $\Delta$ mode cutoff frequency at 9.44~GHz. As a consequence, only for the lowest pump frequencies considered in this work ($\omega_P/(2\pi) \sim 2.2$~GHz) can we detect the dominant sidebands at $\omega_S=\omega_{B_0}$ and $\omega_I=\omega_{B_{2}}$. We note that this is the regime in which our device exhibits the lowest performance (see Fig.~\ref{fig3}c). For this pump frequency, we can further detect the sidebands at $\omega_{B_1}$ and $\omega_{B_{-4}}$ (we do not distinguish between the negative and positive part of the power spectrum).\\

\begin{figure}
    \centering
    \includegraphics[]{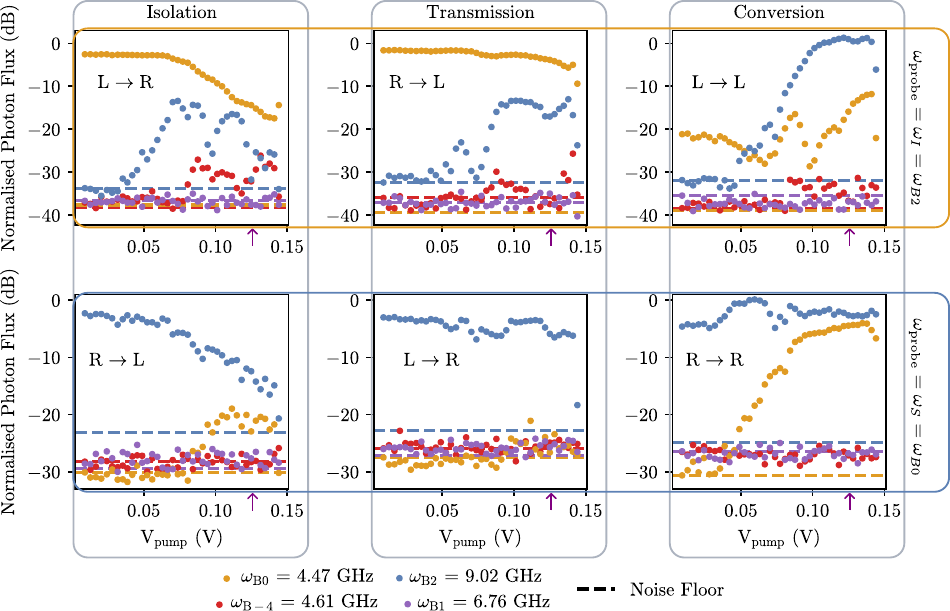}
    \caption{{\bf Power of sideband waves leaking out from one of the device $\Sigma$ port,} in presence of a pump at $\omega_P=2.27$~GHz, as a function of the pump amplitude. We represent the power of  bands lying within the bandwidth of our calibrated power detection setup (output power referenced at fictitious ports addressing each mode of the device, see Sec.~\ref{sec:amplitudecalibration}). Label $i\rightarrow j$ indicate that the probe wave is applied from port $i$ and the generated sideband waves are detected on port $j$, where $i$ and $j$ can be $L$ (left) or $R$ (right). On top panels, the probe frequency is set at $\omega_S=\omega_{B_{0}}$ (respecting the phase-matching condition~\eqref{eq:signal-freq} when the probe is applied from the left port), while on bottom panels, it is set at $\omega_I=\omega_{B_{2}}$ (respecting the phase-matching condition when the probe is applied from the right port). In the leftmost column, we analyze the transmitted spectrum in the direction of isolation of the circulator. As expected the transmitted probe power decreases with pump amplitude (at $\omega_{B_{0}}$ for the top panel, at $\omega_{B_{2}}$ for the bottom panel). The increasing power of band 2 in the top panel may be attributed to secondary reflection of the idler wave after its expected reflection and frequency conversion by the circulator.  In the middle column, we analyze the transmitted spectrum in the direction along which the circulator should not attenuate the probe wave. The decrease in transmitted probe power versus pump amplitude is attributed to internal reflections of the pump wave (see Sec.~\ref{sec:amplitudecalibration}).  In the rightmost column, we detect the reflected field, whose spectrum should reveal frequency conversion of the probe wave. The increase of the power at $\omega_{B_{2}}$ versus pump amplitude in the top panel (at $\omega_{B_{0}}$ for the bottom panel) confirms this conversion and the circulation of waves in the device. The large constant power at $\omega_{B_{0}}$ for the top panel (at $\omega_{B_{2}}$ for the bottom panel) indicates that the probe is partly reflected by the defective junction, without frequency conversion. }
    \label{fig:conversion-grid}
\end{figure}

In Fig~\ref{fig:conversion-grid}, we represent the power of each sideband transmitted and reflected, for a probe wave applied from either port, either at $\omega_S=\omega_{B_0}$ (top panels) or at  $\omega_I=\omega_{B_{2}}$ (bottom panels). In all configurations, the signal and idler bands (respectively in orange and blue) have the largest power as expected. Moreover, the power of the transmitted signal decreases and the power of the reflected idler increases when probing from the left at $\omega_S$ (top left and top right panels), while the power of the transmitted idler decreases and the power of the reflected signal increases when probing from the right at $\omega_I$ (bottom left and bottom right panels). Nevertheless, in both cases, a significant fraction of the probe wave power or of the converted wave power may exit from the wrong port owing to internal reflection by the defective junction. The power of other sidebands remains below  the noise floor of our detection setup for pump amplitudes below critical value. When probing the system from the left port (top line),  the sideband $n=-4$ displays a detectable power above this critical value, which may indicate that this non phase-matched process is involved in the collapse of the device transmission observed over a wide band (see Fig.~\ref{fig:extended-data}).\\

\begin{figure}[h]
    \centering
    \includegraphics[]{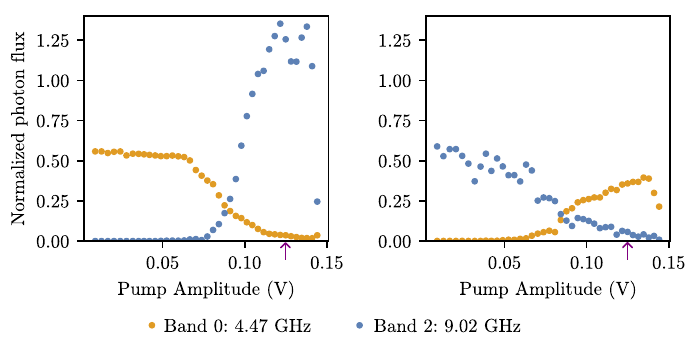}
    \caption{{\bf Conversion measurement by spectrum analysis.} Detected photon flux, normalized by the input flux, for the signal and idler bands as a function of the pump amplitude. The pump is applied from the right $\Delta$ port at 2.27 GHz and at critical amplitude. {\bf Left,} Signal to idler conversion. The signal is sent from the left port. The transmitted signal, or band 0, is measured at the right port and the converted signal, or band 2, is measured at the left port. The transmitted photon flux  decreases while the reflected photon flux  increases. The transmitted photon flux going over 1 indicates that the attenuation of the output line connecting to the right port  is over-estimated at $\omega_I$ or that the attenuation of the input line connecting to the left port is under-estimated at $\omega_S$. {\bf Right,} Idler to signal conversion. The idler tone is sent from the right port. The transmitted idler, or band 2, is measured at the left port and the converted signal, or band 0, is measured at the right port. Again, the transmitted photon flux  decreases while the reflected photon flux  increases, and the photon flux is approximately conserved.}
    \label{fig:conversion-lin}
\end{figure}

In principle, waves circulate within the device while preserving the total number of photons. To investigate further this point, we plot in Fig~\ref{fig:conversion-lin} the photon flux in the transmitted probe wave and the photon flux in the reflected and converted probe wave.  Photon flux conservation is not quantitatively verified and the reflected photon flux is actually larger than the applied photon flux when probing the device from the left $\Sigma$ port at $\omega_S$. We attribute this discrepancy to the finite precision of our calibration method for the attenuation of the system output lines, limited to approximately 3~dB (see Sec.~\ref{sec:amplitudecalibration}). As a consequence, we cannot estimate quantitatively the circulation efficiency of our device. 

\subsection{Pump harmonics}
\label{sec:pump-harmonics}
Using the same calibration of output lines, we now turn to characterizing leakage of the pump and of its  harmonics  onto the $\Sigma$ mode. Here, no probe is applied on the $\Sigma$ mode, the output power is detected at $m\omega_P$ with $m\in \mathbb{N}$ and is referenced  at fictitious ports directly addressing the device.\\

\begin{figure}[h]
    \centering
    \includegraphics[]{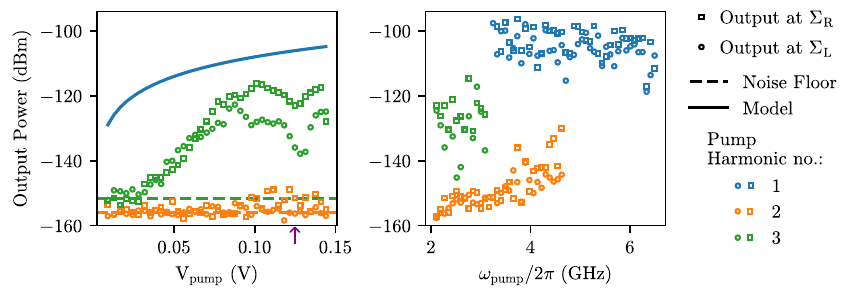}
    \caption{{\bf Power of the pump fundamental tone and harmonics leaking to the $\Sigma$ mode. Left panel:} Power detected  for a pump at $\omega_{P}/2\pi=2.27$~GHz, as a function of pump amplitude. The power is calibrated at fictitious $\Sigma$ ports directly connected to the device (see Sec.~\ref{sec:amplitudecalibration}).  The 3rd harmonic (green) is detectable at all but the lowest pump amplitudes (dashed horizontal lines indicate the detection noise floor). The 2nd harmonic (orange) cannot be generated by four wave mixing, hence why it is negligible except for a few outlying points at high pump amplitude. The pump fundamental tone cannot be directly detected and its power is instead estimated from the input line calibration and the predictions of our single open-junction model.  The purple  arrow indicates the critical pump amplitude at this frequency. {\bf Right panel:} Power of the fundamental pump tone and first harmonics at critical amplitude versus pump frequency. The power of the harmonic no 2 is below the noise floor of our detection setup.}
    \label{fig:pump-harmos}
\end{figure}

In Fig.~\ref{fig:pump-harmos}a, we report the detected power leaking from each $\Sigma$  port, for the second and third harmonic of a pump at $\omega_P/(2\pi)=2.27~$GHz. This frequency is chosen such that the third harmonic lies within the bandwidth of our detection setup and of our calibration. However, we are not able to directly measure the power in the fundamental tone   at such a low frequency. We instead estimate it from our open-junction model (see Fig.~\ref{fig:waveprofiles}, the pump wave leaks to the $\Sigma$ mode in both directions at the -7~dB level). At low pump amplitude, the power of the third harmonic is below our detection threshold. It becomes detectable at higher pump amplitude, increasing with a scaling matching the expected cubic power law, until approaching the critical pump amplitude when it starts to decrease.  The power of the second harmonic remains below detection threshold---except at the highest pump amplitudes for which we detect a weak signal---as expected from our model which only allows the mixing of even numbers of waves. In Fig.~\ref{fig:pump-harmos}b, we report the detected power of the fundamental, second and third  harmonics, when within our detection bandwidth, for a pump of varying frequency and amplitude adjusted at critical value.  We observe the same pattern as  described for a pump at 2.27~GHz.

\subsection{Coherent detection}
\begin{figure}[h]
    \centering
    \includegraphics[width=0.45\columnwidth]{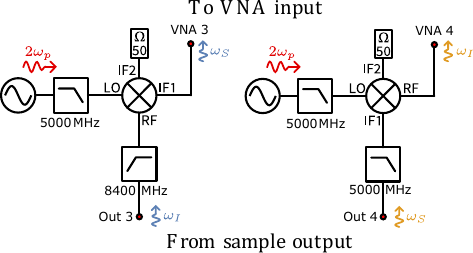}
    \caption{{\bf Up and down-conversion additional setup.} The point Out 3 of the left down-conversion block is inserted at the VNA 3 point in Fig~\ref{fig:setup} and the vector network analyzer is connected at the new VNA 3 point, allowing detection of waves converted to idler frequency at this output. Similarly the point Out 4 of the right up-conversion block is inserted at the VNA 4 point in Fig~\ref{fig:setup}, allowing detection of waves converted to signal frequency. }
    \label{fig:mixer-setup}
\end{figure}

Spectrum analysis does not provide information about the coherence of wave conversion in the circulation process. To investigate this, we modify our setup so that we can  coherently detect converted waves with respect to the sent waves. We add down and up-conversion blocks to the left and right output lines (see Fig~\ref{fig:mixer-setup}) so that reflected and frequency converted waves can be measured with a vector network analyzer. The mixers used are model XM-B2V1-0409C. Again our finite detection bandwidth and the mixers' specifications constrain the range of working points at which this coherent detection may be performed. Thus, we measure the device while applying a pump frequency of 2.27~GHz. The LO frequency is therefore at twice the pump frequency, 4.38~GHz, and generated by a source, model APUASYN20-4 from Anapico. The signal gap is around 4.4~GHz, in the IF range of the mixer, and the idler gap is at around 8.8~GHz, in the RF range. \\

\begin{figure}[h]
    \centering
    \includegraphics{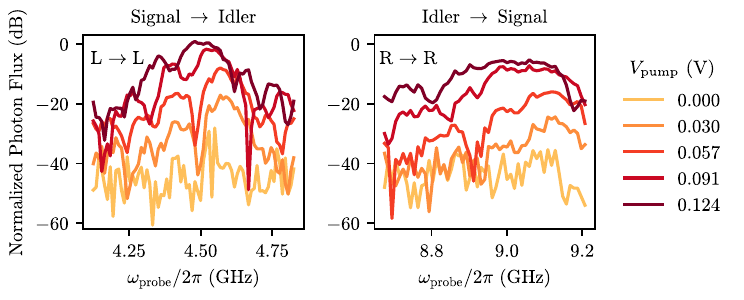}
    \caption{{\bf Coherent conversion detection.} A pump is sent from the right $\Delta$ port at $\omega_P/2\pi=2.27$ GHz. The left and right output lines of the $\Sigma$ mode are routed to a down-conversion and a up-conversion mixer setup, respectively, and then connected to the VNA (see Fig.~\ref{fig:mixer-setup}). On the left (right) panel, a probe signal sent from the left (right) port at the signal (idler) frequency of the circulating process is converted to idler (signal) frequency, and is detected coherently after down-conversion (up-conversion). The colors correspond to varying pump amplitudes, showing stronger conversion with increasing pump amplitude. }
    \label{fig:mixer-reflec}
\end{figure}

Using this updated setup, we measure the conversion of the main circulating process in the relevant frequency bands (Fig~\ref{fig:mixer-reflec}). Signal to idler conversion is obtained from the reflection at the left $\Sigma$ port of the sample, and idler to signal one from the reflection at the right $\Sigma$ port of the sample. We repeat this for varying pump amplitudes. In both cases, the waves are detected against a dark background (noise floor around -90 dB) and its strength scales with the pump amplitude, confirming the coherence of the process. Note that here, we did not calibrate the attenuation of feedlines over the whole probed range. Instead, we use the calibration performed either  at the signal or idler frequency (frequencies for which the circulation process is phase-matched in the low pump amplitude regime) to coarsely correct for feedlines' attenuation over a $\sim600$~MHz band centered around each of these two frequencies.

\section{Fabrication recipe}
\label{sec:fab}
In this section, we describe step by step the process by which our sample was fabricated.
\begin{enumerate}
    \item Base niobium layer
    \begin{itemize}
        \item We sputter 100 nm of Nb on a 2-inch intrinsic silicon wafer, with a 275 µm thickness and a resistivity larger than 10 k.cm.
        \item We spin Microposit S1805 resist and pattern the base layer using optical laser lithography. We develop for 1min in Microposit MF-319. 
        \item We etch the niobium in a Reactive Ion Etching (RIE) machine, using a SF-6 plasma, and taking great care to stop the process as soon as the exposed metal is fully removed as the silicon has a high etching rate. A steep step resulting from excessive etching could be an issue during the deposition of the final aluminum layer, preventing proper film continuity. 
        \item We clean the wafer in a 50°C acetone, and ultrasounds for 2min, rinse in IPA for 30s, and end with a 10s O2 plasma cleaning in the RIE machine. 
    \end{itemize}
    \item Alumina dielectric layer
    \begin{itemize}
        \item We then deposit 80nm of alumina dielectric in a thermal Atomic Layer Deposition machine, the reactants being trimethylaluminum and water. 
        \item We spin Microposit S1813 resist and pattern the mask for the dielectric etching using laser lithography. We develop for 1min in Microposit MF-319.
        \item We etch the alumina in the RIE machine, using a CHF3 plasma.
        \item We clean the wafer in a 50°C acetone, and ultrasounds for 2min, rinse in IPA for 30s, and end with a 10s O2 plasma cleaning in the RIE machine. 
    \end{itemize}
    \item Top aluminum layer
    \begin{itemize}
        \item We coat the wafer with a bilayer of methyl methacrylate-polymethyl methacrylate (MMA EL10-PMMA A6), MMA is baked for 2min at 185°C, PMMA for 30min at 185°C, in prevision of electronic lithography. The electron beam patterning will be done at chip scale, so we coat the wafer an additional unbaked PMMA layer for protection, and then cut it into chips using a dicing saw.
        \item We clean individual chips in IPA, dissolving only the unbaked PMMA. We use ultrasounds for 5min.
        \item We pattern the Dolan bridge structure, as well as large opening for the top plate of the capacitors, using a 20keV electron beam. Note that the capacitors openings are slightly reduced compared to their desired size to compensate for the additional shadow of the later angle evaporation.
        \item We develop the patterned chip in a 3:1 volume IPA:water mix at 6°C for 1min30s, and we rinse it in IPA. We perform a 10s O2 plasma cleaning in the RIE machine to clean the resist residues.
        \item We place the chip inside the vacuum chamber of the electron-beam evaporator. We ion-mill the surface at double angles +30°/-30°, using an argon ion gun, to prepare it for good electrical contact (7sccm flow, $V_{\textrm{discharge}}=40$~V, $V_{\textrm{beam}}=500$~V, $V_{\textrm{acc}}=100$~V, $I=35$~mA, $22$~s duration for each angle). We deposit 35nm of aluminum at +30° angle, then oxidize the surface of the metal in a 3mbar O2 atmosphere for 10 min (note that this low oxydation pressure explains the unusually high plasma frequency of our junctions). We finally deposit 100nm of aluminum at a -30° angle.
        \item We lift-off in a 50°C acetone bath for 1h, sonicate for 2min in a clean 50°C acetone bath and rinse in IPA for 30s.
    \end{itemize}
\end{enumerate}


\makeatother

\end{document}